\documentclass[10pt, compsoc]{IEEEtran}
\makeatletter
\def\ps@headings{%
\def\@oddhead{\mbox{}\scriptsize\rightmark \hfil \thepage}%
\def\@evenhead{\scriptsize\thepage \hfil \leftmark\mbox{}}
\def\@oddfoot{}%
\def\@evenfoot{}}
\makeatother
\usepackage{cite}

\ifCLASSINFOpdf
  \usepackage[pdftex]{graphicx}
  \DeclareGraphicsExtensions{.pdf,.jpeg,.png}
\else
  \usepackage[dvips]{graphicx}
\fi

\usepackage[cmex10]{amsmath}
\usepackage{amssymb, amsthm}
\usepackage{verbatim}

\usepackage{url}

\usepackage{leading}
\leading{12pt}

\usepackage[multiple]{footmisc}

\usepackage{ifthen}

\DeclareMathOperator*{\argmax}{arg\,max}
\DeclareMathOperator*{\argmin}{arg\,min}

\makeatletter
\def\blfootnote{\xdef\@thefnmark{}\@footnotetext}
\makeatother

\begin{document}

\newtheorem{theorem}{Theorem}[section]
\newtheorem{lemma}{Lemma}[section]
\newtheorem{corollary}{Corollary}[section]
\newtheorem{proposition}{Proposition}[section]
\newtheorem{definition}{Definition}[section]
\newtheorem{claim}{Claim}[section]
\newtheorem{remark}{Remark}[section]
\newtheorem{example}{Example}[section]
\renewcommand{\thetheorem}{\arabic{section}.\arabic{theorem}}
\renewcommand{\thelemma}{\arabic{section}.\arabic{lemma}}
\renewcommand{\thecorollary}{\arabic{section}.\arabic{corollary}}
\renewcommand{\theproposition}{\arabic{section}.\arabic{proposition}}
\renewcommand{\theclaim}{\arabic{section}.\arabic{claim}}
\renewcommand{\thedefinition}{\arabic{section}.\arabic{definition}}
\renewcommand{\theexample}{\arabic{section}.\arabic{example}}

\def\bydef{\buildrel \triangle \over =}
\newcommand{\ignore}[1]{}
\newcommand{\bff}{{\bf f}}
\newcommand{\bfi}{{\bf f^i}}
\newcommand{\bfl}{{\bf f_l}}
\newcommand{\bJ}{{\bf J}}
\newcommand{\cG}{{\cal G}}
\newcommand{\cJ}{{\cal J}}
\newcommand{\cN}{{\cal N}}
\newcommand{\cI}{{\cal I}}
\newcommand{\cL}{{\cal L}}
\newcommand{\cV}{{\cal V}}
\newcommand{\cF}{{\cal F}}
\newcommand{\cS}{{\cal S}}
\newcommand{\cP}{{\cal P}}
\newcommand{\hfl}{f_{\hat{l}}}
\newcommand{\hfli}{f_{\hat{l}} ^i}
\newcommand{\hflj}{f_{\hat{l}} ^j}
\newcommand{\fhl}{\hat{f_l}}
\newcommand{\fhli}{\hat{f_l} ^i}
\newcommand{\fhlj}{\hat{f_l} ^j}
\newcommand{\fl}{f_l}
\newcommand{\fli}{f_l ^i}
\newcommand{\flj}{f_l ^j}
\newcommand{\fpi}{f_p^i}
\newcommand{\ftl}{\tilde{f_l}}
\newcommand{\ftli}{\tilde{f_l} ^i}
\newcommand{\fuv}{f_{uv}}
\newcommand{\fuvi}{f_{uv} ^i}
\newcommand{\fuvj}{f_{uv} ^j}
\newcommand{\fvui}{f_{vu} ^i}
\newcommand{\fvuj}{f_{vu} ^j}
\newcommand{\hl}{\hat{l}}
\newcommand{\hbf}{\hat{\mbox{\bf f}}}
\newcommand{\hli}{\hat{\lambda}^i}
\newcommand{\la}{\lambda}
\newcommand{\li}{\lambda^i}
\newcommand{\hlj}{\hat{\lambda}^j}
\newcommand{\lj}{\lambda^j}

\newcommand{\Jsys}{J_{sys}}
\newcommand{\Jli}{J_l ^i}
\newcommand{\Khl}{K_{\hat l}}
\newcommand{\Khli}{K_{\hat l} ^i}
\newcommand{\Khlj}{K_{\hat l} ^j}
\newcommand{\Kl}{K_l}
\newcommand{\Kli}{K_l ^i}
\newcommand{\Klj}{K_l ^j}
\newcommand{\Te}{T_e}
\newcommand{\Tl}{T_l}
\newcommand{\Thl}{T_{\hat{l}}}
\newcommand{\cW}{\mathcal{W}}

\newcommand{\linprog}[6]{
    \begin{alignat}{5}
          \label{#6-1}
          \min        & \quad #1 & \\
          \label{#6-2}
          \text{s.t.} & \quad #2 &, & \quad #3\\
          \label{#6-3}
                      & \quad #4 &, & \quad #5
    \end{alignat}}

\title{Coalitions in Routing Games: \\ A Worst-Case Perspective}

\author{Gideon Blocq and Ariel Orda 
\thanks{G. Blocq and A. Orda are with the Department of Electrical
Engineering, Technion, Haifa 32000, Israel (e-mails: gideon@tx.technion.ac.il,
ariel@ee.technion.ac.il)}}

\IEEEtitleabstractindextext{
	\begin{abstract}
	We investigate a routing game that allows for the creation of coalitions, within the framework of cooperative game theory. 
	Specifically, we describe the cost of each coalition as its maximin value.
	This represents the performance that the coalition can guarantee itself, under any (including worst) conditions.
	We then 
	investigate 
	fundamental solution concepts of the considered cooperative game, namely 
	the {\em core} 
	and a variant of the 
	{\em min-max fair nucleolus}. 
	
	We consider two types of routing games based on the 
	agents' Performance Objectives, namely 
	{\em bottleneck routing games} and {\em additive routing
		games}.
	For bottleneck games we establish that the core includes
	all system-optimal flow profiles and that the nucleolus is 
	system-optimal or disadvantageous for the smallest agent in the system. 
	Moreover, we describe an interesting set of scenarios for which the nucleolus is always system-optimal.
	For additive games, we focus on the 
	fundamental 
	load balancing game
	of routing over parallel links. 
	We establish that, in contrary to bottleneck games, not all system-optimal flow profiles lie in the core.
	However, we describe a specific system-optimal flow profile that does lie in the core and, under assumptions of symmetry, is equal to the nucleolus.
\end{abstract}

}

\maketitle

\thispagestyle{headings}
\pagestyle{headings}


\section{Introduction}
\label{sec:intro}
To date, game theoretic models have been employed in virtually all
networking contexts. These include control tasks at the network layer,
such as flow control and 
routing (e.g.,~\cite{ORS93, altman_2000, La02, Roughgarden:2002}).
In particular, research until now 
in routing games
has mainly focused on {\em non-cooperative} networking games, where the {\em selfish} decision makers (i.e., the 
{\em users} or {\em agents})
cannot communicate and reach a binding agreement on the way they would share the network infrastructure. Moreover,
the main dynamics that were considered were {\em Best-Response}, i.e., each 
agent would observe the present state of the
network and react to it in a self-optimizing manner. ly, the operating points of such systems
were taken to be some equilibria of the underlying non-cooperative game, most notably  Nash equilibria.
Such equilibria are inherently inefficient
and, in
general, exhibit suboptimal network performance. As a result,
the question of how bad the quality of a 
non-cooperative equilibrium is with respect to a centrally enforced optimum has
received considerable attention e.g.,~\cite{KoutsoupiasP09,Roughgarden:2002}. 

However, there is a growing number of networking scenarios where, while there is competition among 
self-optimizing agents, 
there is also a possibility for these
agents to communicate, negotiate and {\em reach a binding agreement}
(see, e.g., \cite{YaicheMR00, Han:2009, Gibbens:2008, LXWG2011, SaadHDHB09, AntoniouKJPS09, Tembine:2012, Singh:2012, Saad_coalitionalgame}).
Indeed, in many scenarios,
the competition is among business organizations, which can, and often do, reach agreements (e.g., SLAs) on the way
that they provide, consume or share the network resources. The proper framework for analyzing such
settings is that of {\em Cooperative Game Theory} \cite{Saad_coalitionalgame, Myerson}.  
Such a paradigm transfer, from non-cooperative 
to cooperative
games, calls to revisit fundamental concepts. Indeed, the operating point of the network is not an equilibrium of a non-cooperative game anymore, but rather a solution of a cooperative
game. Accordingly, the performance degradation of such systems should be considered at new operating
points.
In the realm of routing games, such an operating point has been proposed in \cite{BO12}, which considered the adoption of the Nash Bargaining Scheme (NBS) \cite{JNashNBS} as a way of reducing the potentially high inefficiency of the Nash Equilibrium.\footnote{The NBS has been considered in other networking scenarios, e.g., \cite{YaicheMR00}. However, to the best of our knowledge, \cite{BO12} is the first to consider the NBS in the context of routing games.}
Nevertheless, the NBS only contemplates two scenarios, namely the ``grand coalition'' (i.e., an agreement reached by all agents) and the {\em disagreement point}, i.e., the outcome of the fully non-cooperative scenario. Thus, while bargaining between entities is encouraged 
at the NBS, it might be advantageous for groups to deviate from the agreed strategy and form {\em subcoalitions}. 
Such deviations are to be avoided, since it returns the network to its inefficient non-cooperative scenario. 
Accordingly, we focus on routing strategies 
for which deviations of subcoalitions do not occur when agents behave in a rational manner. A design guideline would be to have a mediator, e.g., a network administrator, propose to all agents in the network only to route their flow according to such stable operating points.
More importantly, this gives a credible guarantee to all agents that the agreed upon solution will be implemented.
To represent the set of stable operating points, 
we focus on the {\em core}\cite{Myerson} of our game.
Since the core might include several solutions, we further consider a specific
solution
in the core with min-max fairness properties, which is a variant of the {\em nucleolus} \cite{Schmeidler_Nuc}. 


When considering a coalitional game with $N$ agents, 
a major question is what cost should be attributed to each of the $2^N$ possible coalitions.
In other words, which scenario can each coalition expect in the network when deviating from the agreed upon solution?
In the traditional game theoretic setting all agents try to minimize their cost function, however in the 
context of networking, it is often plausible to contemplate scenarios 
in which some agents do not care about optimizing their own cost, but seem to act maliciously towards others \cite{Moscibroda:2006, Karakostas:2007, Chakrabarty, KarakostasV07, BabaioffKP09, altman:hal-01249188}. 
Such behavior is due to a range of reasons, e.g., hackers or rivaling companies that 
aim to degrade network quality. 
Moreover and perhaps more notably, 
it may happen that some agents are not aware of how to optimize their cost, hence they might exhibit seemingly ``irrational'', thus unexpected, behavior.
In light of such settings, we model the cost of a coalition to equal its {\em worst-case} scenario, in order to investigate what amount of resources it can guarantee under any (including worst) conditions.
Specifically, we describe the cost of each coalition as its maximin value \cite{MaschlerGT}, i.e., as the corresponding best-response to the {\em maximin} strategy of the agents outside the coalition.\footnote{When dealing with payoffs instead of costs, this corresponds to the minimax value of a coalition \cite{MaschlerGT}.
Alternatively, this scenario can be viewed as a Stackelberg game \cite{Myerson}, where an adversary, which acts as ``leader", tries to maximize the cost of the coalition, which acts as ``follower". Recently, in \cite{altman:hal-01249188} the maximin value of two-player games has been studied in the restricted setting of users with $M/M/1$ cost functions.}\footnote{In our considered game, the maximin strategy of the agents outside the coalition does not necessarily equal the minimax strategy of the coalition.}
This represents the cost that a coalition can guarantee itself, even under the pessimistic expectation that the agents outside the coalition have unpredictable or even malicious objectives. 
More importantly, this way of modeling gives an {\em a priori} insight into any agreement proposed by a mediator. Clearly, no agreement can be reached between all agents if a subset of agents can guarantee itself a better outcome at its worst-case 
scenario.

We concretize our study of coalitions in routing games 
by considering 
two types of 
games based on the structure of the agents' performance objectives. The first game considers agents with {\em bottleneck objectives},
i.e., their performance is determined by the worst component (link) in the network \cite{BannerO07,BuschM09,  ColeDR12, HarksKM13}. 
{\em Bottleneck routing games} have been shown to emerge in many practical scenarios. For example, in wireless networks, the weakest link in a transmission is determined by the node with the least remaining battery power. Hence, each agent would route traffic so as to maximize the smallest battery lifetime along its routing topology.
Additionally, bottleneck routing games arise in congested networks where it is desirable to move traffic away from congested hot spots.
For further discussion and additional examples see \cite{BannerO07}.
The second type of game considers 
agents with {\em additive} performance measures, where the performance is determined by the sum of its link performances,
e.g., delay or packet loss. Much of the current literature on networking games has focused on such games, e.g., \cite{ORS93, KoutsoupiasP09,altman_2000, La02, Roughgarden04, Roughgarden:2002, Nisan:2007, Harks11, rough_schopp11, Wan12}, albeit in the traditional setting of non-cooperative agents. 

\subsection*{Our Contribution}
We focus our study on the atomic splittable routing model \cite{ORS93, BannerO07}, in which 
each agent sends its non-negligible demand to the destination by splitting it over a set of paths in the network. Moreover, agents are able to cooperate and form coalitions. 
In particular, we formulate a Non-Transferable Utility Coalitional Game \cite{Myerson} and describe the cost of each coalition as its maximin value.
For bottleneck routing games we establish that coalitions with larger aggregated demand receive a smaller cost at their maximin flow profile. 
Through this result, we describe a set of 
flow profiles
that are stable\footnote{Formally, the core \cite{Myerson} considers a set of stable cost allocations, each of which, in our routing game, refers to a set of flow profiles. While abiding by the mathematical definition of the core, in the context of this paper, we will also refer to the corresponding routing strategy flow profiles as being stable.}, i.e., from which it is not worthwhile for any coalition to deviate. 
In particular, to represent the set of stable solutions, we focus on the {\em core} and a variant of the {\em min-max fair nucleolus}. 
For bottleneck games we establish that any system-optimal flow
profile lies in the core. Moreover, at the nucleolus we establish that
\textbf{(1)} all agents route their flow according to the system optimum or \textbf{(2)} only the smallest agent experiences performance that is worse
than at the system optimum. 
Next, we describe an interesting set of scenarios in which the nucleolus is always system-optimal.
Specifically, 
and counter-intuitively, 
we establish that in a network where the two smallest agents are of equal size, the nucleolus is always system-optimal. A special case of this scenario is when all agents have equal demands.

For additive routing games we focus on the framework of routing in a ``parallel links'' network. 
Beyond being a basic framework of routing, this is the {\em generic 
framework of load balancing} among servers in a network.  It has been 
the subject of numerous studies in the context of non-cooperative networking games, e.g.,
~\cite{ORS93, KoutsoupiasP09, Roughgarden04, La02, Harks11, Wan12},
to name a few. 
Quite surprisingly, we establish that 
at the maximin strategy, the ``malicious'' agents outside the coalition act as if they were a continuum of infinitesimal (i.e., {\em nonatomic}) self-optimizing agents. 
With the above structural result at hand, we establish that, in contrast to bottleneck games, not all system-optimal flow profiles are necessarily stable. 
Nevertheless, we prove that
a particular system-optimal flow profile does lie in the core, namely where all agents send their demand proportionally with respect to the system optimum.
Finally, we show that when agents have equal demands, this proportional system-optimal flow profile is also equal to the
{\em nucleolus}.
\section{Model and Game Theoretic Formulations}
\label{sec:model}
\subsection{Model}
\begin{figure}[h!]
  \centering
    \includegraphics[width=0.5\textwidth]{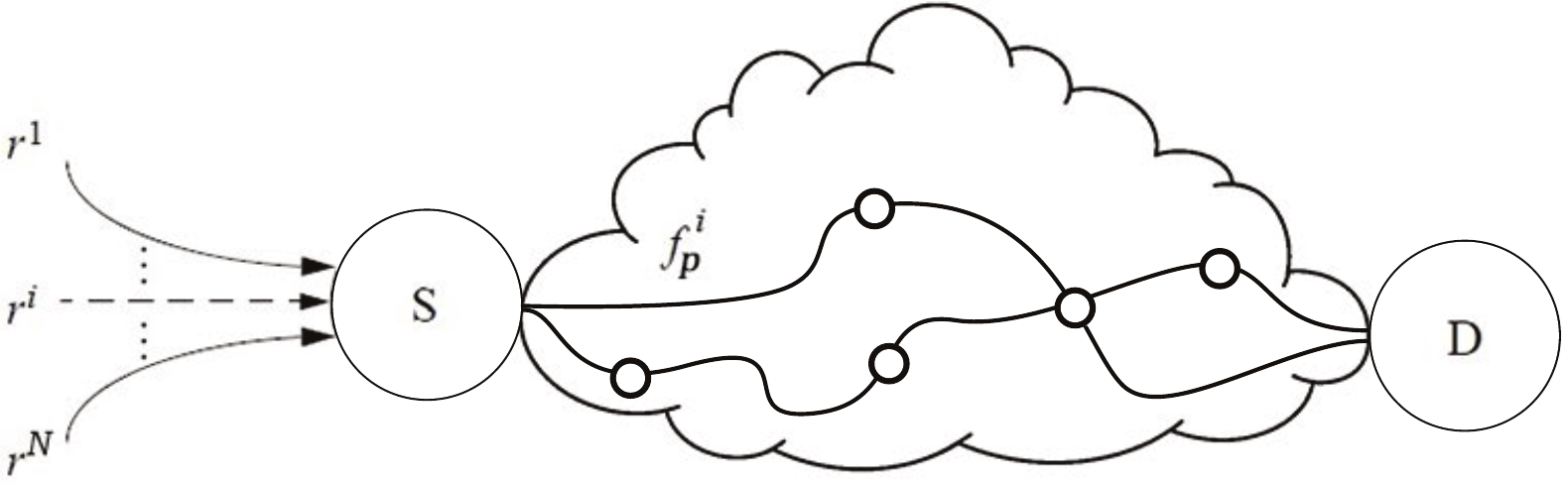}
  \caption{Communication Network}
\label{fig:general_graph}
\end{figure}

We consider a 
set $\cN = \{ 1,2, \ldots , N \}$ of
selfish ``users'' (or, ``players'', ``agents''), which share a communication network modeled by a directed graph $G(V,E)$, as illustrated in Figure \ref{fig:general_graph}.
We denote by $\cP$ the set of all paths in the network.
Each user  $i \in \cN$ has a traffic demand $r^i$ and all users share a common source $S$ and common destination $D$. Denote the aggregated demand of users in the coalition $S \subseteq \cN$ as $r^S$, and the demand of all the users as $R$, i.e., $R = \sum_{i \in \cN} r^i$. 
A user ships its demand from $S$ to $D$ by splitting it along the paths in
$\cP$, i.e., user $i$ decides what fraction of
$r^i$ should be sent on 
each path. 
We denote by $\fpi$, the flow 
that user $i \in \cN$ sends on path $p \in \cP$. User $i$ can fix any
value for $\fpi$, as long as $\fpi \geq 0$ (non-negativity constraint)
and $\sum_{p \in \cP} \fpi = r^i$ (demand constraint);
this assignment of traffic to paths, $\mathbf{f^i} = \{f_p^i\}_{p \in \cP}$ shall also be referred to as the {\em routing strategy} of user $i$. 
The {\it (routing strategy) flow profile} $\bff$ is
the vector of all user routing strategies, 
$\bff = ( {\bf f^1} , {\bf f^2} , \ldots , {\bf f^N} )$. 
We say that a user's 
routing strategy is {\it feasible} if its components obey the nonnegativity
and demand constraints.
We say that a 
flow profile $\bff$ is feasible if
it is composed of feasible routing strategies and we denote by
$\bf{F^i}$ the convex and compact set of all feasible $\bf f^i$'s.
Also denote 
the set $\bf{F}$ as the set of all feasible flow 
profiles. 
Turning our attention to a path $p \in \cP$, let $f_p$ be the total flow
on that path i.e., $f_p = \sum_{i \in \cN} f_p^i$; also denote by $f_e^i$ the flow that $i$ sends on link $e \in E$, i.e., $f_e^i = \sum_{p|e \in p}f_p^i$. Similarly, the total flow on link $e \in E$ is denoted by $f_e = \sum_{i \in \cN} f_e^i$.
For any coalition of users $S \subseteq \cN$ the aggregated flow on path $p$ and link $e$ is denoted by respectively $f^S_p = \sum_{i \in S} f^i_p$ and $f^S_e = \sum_{i \in S} f^i_e$.

\begin{figure}[h!]
	\centering
	\includegraphics[width=0.45\textwidth]{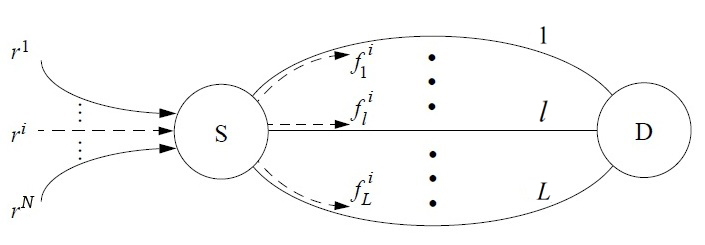}
	\caption{Parallel links network}
	\label{fig:par_links}
\end{figure}

Occasionally, we focus on the framework of routing in a ``parallel links'' network. 
In such cases $G(V,E)$ corresponds to a graph with parallel ``links'' (e.g., communication links, servers, etc.) $\cL = \{ 1,2, \ldots ,L \}$, $L>1$, and a user ships its demand by splitting it over the links $\cL$, see Figure \ref{fig:par_links}.
As observed in \cite{ KorilisLO97}, it constitutes an appropriate model for seemingly unrelated networking problems. For example, in a QoS-supporting network architecture,
bandwidth may be separated among different virtual paths, resulting effectively in a system of parallel and noninterfering “links” between the source and destination. 
Additionally, one can consider a corporation or organization that receives service from a number of different network providers. The corporation can split its total flow over the various network facilities (according to performance and cost considerations), each of which can be represented as a “link” in the parallel link model.
More generally, the problem of routing over parallel links is, essentially, the generic problem of load balancing among several servers, and it has been the  subject of numerous studies, including the seminal paper 
\cite{KoutsoupiasP09}
and many others, e.g., \cite{ORS93, Roughgarden04, La02, Harks11, Wan12}.

In $G(V,E)$, we associate with each link a performance function $T_e(\cdot)$, which corresponds to the {\em cost per unit of flow} through link $e$ and only depends on the total flow $f_e$. 
Furthermore, we impose the following assumptions on $T_e(f_e)$:
\begin{list}{S\theenumi}{\usecounter{enumi}}
	\item $T_e ( f_e )$ is continuous and strictly increasing in $f_e$.
	\item $T_e : [0, \infty ) \rightarrow [0, \infty]$.
	\item For every flow profile $\mathbf{f}$, if not all costs
	are finite then at least one user with infinite cost 
	can change its own strategy to make its cost finite.
\end{list}
Cost functions that comply with the above assumptions shall
be referred to as {\em standard}.
The performance measure of a user $i \in \cN$ is given by a cost function $J^i ( \bff )$.
In bottleneck routing games, $J^i ( \bff )$ corresponds to the performance of the worst-case link, and in additive routing games it corresponds to the sum of all link performances in the system. 
An $N$-tuple of positive values $\bJ = (J^1, J^2, \ldots , J^N)$ is said to be
a {\em feasible cost vector} if there is a feasible routing strategy flow profile $\bff \in \bf {F}$ such that, for all $1 \leq i \leq N$,
$J^i = J^i(\bff)$. Denote the set of feasible cost vectors by $\mathcal{\cJ}$.

\subsection{Bottleneck routing cost function}
Following \cite{BannerO07}, we represent the cost of a user $i \in \cN$
as the worst performance of any link in the network that $i$ sends a positive amount of flow on.
Thus, the cost of user $i$ equals
\begin{equation}\label{eq:bottleneck_cost}
J^i ( \bff ) \triangleq \max_{e \in E| f_e^i>0} T_e(f_e).
\end{equation}
We measure the welfare of the system through the cost of the worst performing link in the network, i.e., through a (``social'') cost function $\Jsys$ defined as $J_{sys} \triangleq \max_{e \in E | f_e>0} T_e(f_e)$. For bottleneck routing games we consider users with {\em standard} cost functions.

\subsection{Additive routing cost function}
Another important class of problems is when users are interested additive performance measures, e.g., delay or packet loss. In this case, $\Te$ may correspond to the total delay of link $e$.
Following \cite{ORS93, KorilisLO97, Cominetti:2009} and many others,
we consider users whose  cost functions are of the following form:
\begin{equation}\label{eq:additive_cost}
J^i(\mathbf{f})\triangleq\sum_{p \in \cP} f_p^i \sum_{e \in p} T_e(f_e).
\end{equation}
Moreover, we consider $T_e ( f_e )$ to be convex and continuously differentiable.
As commonly assumed in the literature (e.g., \cite{KorilisLO97, La02, Roughgarden:2002, Roughgarden04, Cominetti:2009, rough_schopp11, Harks11}), for additive routing games 
the system's cost is equal to the sum of the individual costs of the players, i.e.,
$\Jsys = \sum_{i \in \cN} J^i$. 
The system's cost thus equals: 
\begin{equation} \label{eq:additive_system_cost}
\Jsys = \sum_{i \in \cN} \sum_{p \in \cP} \sum_{e \in p} f_p^i T_e(f_e) =
\sum_{e \in E} f_e T_e(f_e).
\end{equation}

\subsection{System Optimization}
We denote the optimal value of the system's cost as $J^*_{sys}$, i.e., the minimal value of $\Jsys$ over all feasible routing strategy flow profiles. Any flow profile that corresponds to the system optimum, we denote by $\bf{f}^*$ $= (f_p^*)_{p \in \cP}$. 
For both bottleneck and additive routing games, 
$\Jsys$ depends only on the total flows on the links. 
Accordingly, $\bf{f}^*$ is an 
optimal vector of link flows, i.e., 
$J^*_{sys} = \max_{e\in E|f^*_e>0} T_e(f^*_e)$ and $J^*_{sys} = \sum_{e \in E} f_e^* T_e (f_e^*)$ for respectively bottleneck and additive routing games.
A recurring flow profile in this study is the {\em proportional flow profile} where all users send their flow proportionally with regard to the system optimum: for any system-optimal flow profile $\bf f^*$, any user $i$ and path $p$, $f_p^i = \frac{r^i}{R}f_p^*$.
We denote the cost vector at the proportional flow profile as the {\em Proportional Cost Allocation (PCA)}. 
Consequently, at the PCA, for each user $i \in \cN$ with bottleneck costs, $J^i(\mathbf{f^*}) = J^*_{sys}$, and for users with additive costs,
\begin{equation}
\label{proportional_alloc}
J^i(\mathbf{f^*}) =  \sum_{p \in \cP} \frac{r^i}{R} f_p^* \sum_{e \in p} T_e(f_e^*) = \frac{r^i}{R}J^*_{sys},~\forall i \in \cN.
\end{equation}
\subsection{Nonatomic users}
In this study we focus on
a finite set of (nonzero-size) users that can split their flow among the links.
Nevertheless, 
some of the following results state that at times a user may behave as if it were a continuum of infinitesimal self-optimizing users, referred to as a set of {\em nonatomic users}. 
A nonatomic user places its demand on a single path $\hat{p}$, for which $\sum_{e \in \hat{p}} T_{e}(f_{e}) = \min_{p \in \cP} \sum_{e \in p} T_e(f_e)$ \cite{WardropJ52}. 
If a finite user $i$ behaves as if it were a set of self-optimizing nonatomic users, it follows 
that, $\forall \hat{p},p\in \cP$:
\begin{equation}
\label{non_atomic_user_conditions}
\text{if}~f^i_{\hat{p}} > 0\rightarrow~ \sum_{e \in \hat{p}}T_e(f_e) \leq \sum_{e \in p} T_e(f_e).
\end{equation}
We will refer to (\ref{non_atomic_user_conditions}) as the {\em best-response behavior of a set of nonatomic users}. 

\subsection{Coalitional game} 
We proceed to formalize our coalitional game, by attributing a set of costs to every coalition $S \subseteq N$. Note that our cost functions represent a variety of costs, e.g., delay, 
which are not considered to be a commodity that users can freely transfer between themselves, 
hence we define a {\em Non-Transferable Utility} (NTU) Game \cite{Myerson}, as follows.
\begin{definition}
\label{def:NTU_game}
A NTU coalitional game consists of a mapping $V(\cdot)$ that assigns to each coalition $S \subseteq \cN$ 
a set of outcomes $V(S)\subseteq \mathbf{R}^{|S|}$, which is non-empty, closed and convex.\footnote{In 
\cite{Myerson} additional conditions are cited, which 
trivially follow for our considered game. 
}
\end{definition}
\noindent A coalition behaves as a single user controlling the flow of its participants and $V(S)$ represents the set of feasible cost allocations that $S$ can achieve for itself.

\section{Bottleneck Routing Games}
\subsection{Maximin flow profile}
As explained in the introduction, our goal is to propose
stable and fair routing strategies 
to all $\cN$ users. However, once a coalition of users $S$ decides to deviate from our proposed solution, we model it to incur a worst-case cost. 
Specifically, we represent the cost of every coalition that deviates (and there are $2^{|\cN|}-1$ such possible coalitions), as its maximin value in the game between $S$ and the users in $\cN \backslash S$. 
The maximin value of $S$ corresponds to the lowest cost $S$ can guarantee against the worst (for $S$) strategy of the users outside the coalition. 
In other words, all users in $\cN \backslash S$ aim to 
maximize 
$J^S$ by sending their flow according to $\bf f^{\boldsymbol{\cN} \backslash S}$. 
Given $\bf f^{\boldsymbol{\cN} \backslash S}$, $S$ behaves as a single user controlling the flow of its participants and aims to minimize their combined cost $J^{S}$.
Clearly, by cooperating, the users in $\cN \backslash S$ can only increase their damage to $S$. 
Therefore, when considering the maximin value of $S$, we represent all users in $\cN \backslash S$ as a single malicious user that aims to maximize the cost of $S$. 
For each coalition we can now view this as a game between two players, $S$ and ${\cN} \backslash S$. 
Given any routing strategy $\mathbf{f^{\boldsymbol{\cN} \backslash S}}$ denote the {\em best-response strategy} of coalition $S$ as 
\begin{equation}
\label{eq:best-response-strategy}
\mathbf{\hat{f}^S}(\mathbf{f^{\boldsymbol{\cN} \backslash S}}) = \argmin_{\bf f^S \in F^S} \{J^S(\mathbf{f^{\boldsymbol{\cN} \backslash S}, f^S}) \}.
\end{equation}
The maximin strategy of the users in ${\cN} \backslash S$ is equal to
\begin{equation}
\label{eq:malicious_maximin_strategy}
\mathbf{\hat{f}^{\boldsymbol{\cN} \backslash S}} =\argmax_{\bf f^{\boldsymbol{\cN} \backslash S} \in F^{\boldsymbol{\cN} \backslash S}} \{\min_{\bf f^S \in F^S}  J^S(\mathbf{f^{\boldsymbol{\cN} \backslash S}, f^S}) \} 
\end{equation}
and given $\mathbf{\hat{f}^{\boldsymbol{\cN} \backslash S}}$, the coalition $S$ send its demand according to its best-response strategy, i.e.,
\begin{equation}
\label{eq:coalition_maximin_strategy}
\mathbf{\hat{f}^S}(\mathbf{\hat{f}^{\boldsymbol{\cN} \backslash S}}) = \argmin_{\bf f^S \in F^S} \{J^S(\mathbf{\hat{f}^{\boldsymbol{\cN} \backslash S}, f^S}) \}.
\end{equation}
We refer to $\mathbf{(\hat{f}^{\boldsymbol{\cN} \backslash S}}, \mathbf{\hat{f}^S}(\mathbf{\hat{f}^{\boldsymbol{\cN} \backslash S}}))$ as the {\em maximin 
(routing strategy) 
flow profile}.\footnote{In \cite{BO07} it has been established that, for bottleneck routing games, $J^S(\mathbf{f^{\boldsymbol{\cN} \backslash S}, f^S})$ is not necessarily continuous in $\bf F^S$. Thus, at the maximin flow profile we will consider the set of flow profiles $(\mathbf{\hat{f}^{\boldsymbol{\cN} \backslash S}, \hat{f}^S})$ such that
\begin{equation*}
J^S(\mathbf{\hat{f}^{\boldsymbol{\cN} \backslash S}, \hat{f}^S(\hat{f}^{\boldsymbol{\cN} \backslash S})}) \geq \sup_{\mathbf{f^{\boldsymbol{\cN} \backslash S} \in F^{\boldsymbol{\cN} \backslash S}}} J^S(\mathbf{f^{\boldsymbol{\cN} \backslash S}, \hat{f}^S(f^{\boldsymbol{\cN} \backslash S})}) - \delta,
\end{equation*}
and given any $\bf f^{\boldsymbol{\cN} \backslash S}$
\begin{equation*} 
J^S(\mathbf{f^{\boldsymbol{\cN} \backslash S}, \hat{f}^S}) \leq \inf_{\mathbf{f^S \in F^S}} J^S(\mathbf{f^{\boldsymbol{\cN} \backslash S}, f^S}) + \epsilon
\end{equation*}
for some small enough $\epsilon,\delta \geq0$. 
Consequently, due to the compactness of $\bf F^S$ and $\bf F^{\boldsymbol{\cN} \backslash S}$, it is clear that a maximin profile always exists, however it is not necessarily unique.
For additive routing games, $J^S(\mathbf{f^{\boldsymbol{\cN} \backslash S}, f^S})$ is a continuous function, thus $\epsilon, \delta = 0$.}
To simplify the notation, we will frequently refer to $\mathbf{\hat{f}^S}(\mathbf{\hat{f}^{\boldsymbol{\cN} \backslash S}})$ as $\mathbf{\hat{f}^S}$.
From the point of view of the coalition, at the maximin flow profile, the users outside of the coalition act in a malicious manner.  
Since a coalition $S$ behaves as a single user controlling the flow of its participants, it follows from (\ref{eq:bottleneck_cost}) that in a bottleneck routing game,
\begin{equation} \label{eq:maximin_bottleneck_cost}
J^S(\mathbf{f^{\boldsymbol{\cN} \backslash S}, f^S}) =  \max_{\{e\in E| f^S_e>0\}} T_e(f^{\cN \backslash S}_e + f^S_e),
\end{equation}
where $f^S_e$ represents the combined flow of all users $i \in S$ on link $e$, i.e., $f^S_e = \sum_{i \in S} f^i_e$.
In order to describe the behavior of a coalition at the maximin flow profile, we first establish the following lemma.  
\begin{lemma}\label{lem:larger_demand_smaller_cost}
Consider a bottleneck routing game. For any two coalitions $S,T \subseteq \cN$ with $r^{T} \geq r^{S}$:
$$J^{T}(\mathbf{\hat{f}^{\boldsymbol{\cN} \backslash T}, \hat{f}^{T}}) \leq J^{S}(\mathbf{\hat{f}^{\boldsymbol{\cN} \backslash S}, \hat{f}^{S}}).$$
\end{lemma}
\begin{proof}
We establish the lemma by constructing a routing strategy $\bf \bar{f}^{\boldsymbol{\cN} \backslash S}$, such that 
\begin{equation}\label{eq:bigger_better}
J^{T}(\mathbf{\hat{f}^{\boldsymbol{\cN} \backslash T}, \hat{f}^{T}}) \leq J^{S}(\mathbf{\bar{f}^{\boldsymbol{\cN} \backslash S}, \hat{f}^{S}(\bar{f}^{\boldsymbol{\cN} \backslash S})}).
\end{equation}
Then, as a result of (\ref{eq:bigger_better}) and (\ref{eq:malicious_maximin_strategy}), the lemma follows. 
For every path $p$, consider $\bar{f}^{\cN \backslash S}_p \triangleq g^{\cN \backslash S}_p + h^{\cN \backslash S}_p$, where $g^{\cN \backslash S}_p = \hat{f}^{\cN \backslash T}_p$. 
Since $\sum_{p \in \cP} \hat{f}^{\cN \backslash T}_p = r^{\cN \backslash T} \leq r^{\cN \backslash S}$, we construct $h^{\cN \backslash S}_p$ by sending the remaining demand (i.e., $r^{\cN \backslash S} - r^{\cN \backslash T}$) on the paths where $\hat{f}^T_p>0$, while abiding by the constraint $h^{\cN \backslash S}_p < \hat{f}^T_p$. This constraint is feasible since $\sum_{p \in \cP} h^{\cN \backslash S}_p = r^{\cN \backslash S} - r^{\cN \backslash T} = r^T-r^S< r^T$.
Denote the best-response strategy $\bf \hat{f}^{S}(\bar{f}^{\boldsymbol{\cN} \backslash S})$ as $\bf \bar{f}^{S}$ and assume by contradiction that 
\begin{equation}\label{eq:bigger_better_2}
J^{T}(\mathbf{\hat{f}^{\boldsymbol{\cN} \backslash T}, \hat{f}^{T}}) > J^{S}(\mathbf{\bar{f}^{\boldsymbol{\cN} \backslash S}, \bar{f}^{S}}).
\end{equation}
From (\ref{eq:bigger_better_2}) it follows that
\begin{align}\label{eq:bigger_better_3}
\max_{e\in E|\hat{f}^T_e>0} T_e(\hat{f}^{\cN \backslash T}_e+\hat{f}^T_e) 
&> \max_{e\in E|\bar{f}^S_e>0} T_e(\bar{f}^{\cN \backslash S}_e+\bar{f}^S_e)
\\ \nonumber
&= \max_{e|\bar{f}^S_e>0} T_e(\hat{f}^{\cN \backslash T}_e+h^{\cN \backslash S}_e + \bar{f}^S_e).
\end{align}
We now define a new routing strategy $\bf f^T$ where for any path $p$, $f^T_p = h^{\cN \backslash S}_p + \bar{f}^S_p$. 
This strategy is feasible since $r^T = \sum_{p \in \cP} f^T_p = \sum_{p \in \cP} h^{\cN \backslash S}_p + \bar{f}^S_p = r^T-r^S+r^S$.
Thus, for any link $e$, $f^T_e = h^{\cN \backslash S}_e + \bar{f}^S_e$. Since $\bf \hat{f}^T$ is the best-response strategy to $\bf \hat{f}^{\boldsymbol{\cN} \backslash T}$, we get that
\begin{align}\label{eq:bigger_better_4}
&\max_{\{e\in E|h^{\cN \backslash S}_e+\bar{f}^S_e>0\}} T_e(\hat{f}^{\cN \backslash T}_e+h^{\cN \backslash S}_e + \bar{f}^S_e)
\geq 
\\ \nonumber
&\max_{\{e\in E|\hat{f}^T_e>0\}} T_e(\hat{f}^{\cN \backslash T}_e+\hat{f}^T_e).
\end{align}
As a result of (\ref{eq:bigger_better_3}) and (\ref{eq:bigger_better_4}) 
it follows that
\begin{align}\label{eq:bigger_better_5}
&\max_{\{e\in E|h^{\cN \backslash S}_e+\bar{f}^S_e>0\}} T_e(\hat{f}^{\cN \backslash T}_e+h^{\cN \backslash S}_e + \bar{f}^S_e)
>
\\ \nonumber
&\max_{\{e\in E|\bar{f}^S_e>0\}} T_e(\hat{f}^{\cN \backslash T}_e+h^{\cN \backslash S}_e + \bar{f}^S_e).
\end{align}
Therefore,
\begin{align}\label{eq:bigger_better_6}
&\max_{\{e\in E|h^{\cN \backslash S}_e+\bar{f}^S_e>0\}} T_e(\hat{f}^{\cN \backslash T}_e+h^{\cN \backslash S}_e + \bar{f}^S_e)
=
\\ \nonumber
&\max_{\{e\in E|h^{\cN \backslash S}_e > 0, \bar{f}^S_e=0\}} T_e(\hat{f}^{\cN \backslash T}_e+h^{\cN \backslash S}_e).
\end{align}
By definition, for any link $e$ where $h^{\cN \backslash S}_e>0$ it holds that $\hat{f}^T_e>0$. Thus, from (\ref{eq:bigger_better_5}) and (\ref{eq:bigger_better_6}) it follows that
\begin{align}\label{eq:bigger_better_7}
&\max_{\{e\in E|\hat{f}^T_e > 0\}} T_e(\hat{f}^{\cN \backslash T}_e+h^{\cN \backslash S}_e)
\geq
\\ \nonumber
&\max_{\{e\in E|h^{\cN \backslash S}_e > 0\}} T_e(\hat{f}^{\cN \backslash T}_e+h^{\cN \backslash S}_e)
=
\\ \nonumber
&\max_{\{e\in E|h^{\cN \backslash S}_e > 0, \bar{f}^S_e=0\}} T_e(\hat{f}^{\cN \backslash T}_e+h^{\cN \backslash S}_e)
=
\\ \nonumber 
&\max_{\{e\in E|h^{\cN \backslash S}_e+\bar{f}^S_e>0\}} T_e(\hat{f}^{\cN \backslash T}_e+h^{\cN \backslash S}_e + \bar{f}^S_e).
\end{align}
Finally, from (\ref{eq:bigger_better_7}) and (\ref{eq:bigger_better_4}) we get that
\begin{align*}
&\max_{\{e\in E|\hat{f}^T_e > 0\}} T_e(\hat{f}^{\cN \backslash T}_e+h^{\cN \backslash S}_e)
\geq
\\ \nonumber
&\max_{\{e\in E|\hat{f}^T_e>0\}} T_e(\hat{f}^{\cN \backslash T}_e+\hat{f}^T_e),
\end{align*}
which is a contradiction since $h^{\cN \backslash S}_e < \hat{f}^T_e$ for all $e \in E$.
Thus, (\ref{eq:bigger_better}) follows and the lemma is established.
\end{proof}
According to Lemma \ref{lem:larger_demand_smaller_cost}, coalitions with a larger demand receive a smaller cost at their maximin flow profile. In other words, the threat of the users outside the coalition decreases when the coalition's demand increases.

\subsection{Worst-Case coalitions}
We now proceed to describe our coalitional game as defined in Definition \ref{def:NTU_game}. 
In particular, for each coalition $S$, denote by $\mathbf{(\hat{f}^{\boldsymbol{\cN} \backslash S}, \hat{f}^S)}$ a maximin routing strategy flow profile of the game between $S$ and $\cN \backslash S$, i.e., the flow profile in which $\cN \backslash S$ acts according to (\ref{eq:malicious_maximin_strategy}) and $S$ acts according to (\ref{eq:coalition_maximin_strategy}). 
Also denote $\bf J(\mathbf{\hat{f}^{\boldsymbol{\cN} \backslash S}, \hat{f}^S}) \in \cJ$ as the cost vector of all users, when sending their demand according to $\mathbf{(\hat{f}^{\boldsymbol{\cN} \backslash S}, \hat{f}^S)}$. Furthermore, denote by $\bf J^S(\mathbf{\hat{f}^{\boldsymbol{\cN} \backslash S}, \hat{f}^S})$ the projection of $\bf J(\mathbf{\hat{f}^{\boldsymbol{\cN} \backslash S}, \hat{f}^S})$ onto $\mathbb{R}^S$, i.e., the cost vector of the users in the coalition $S$. 
Note that $J^S(\mathbf{\hat{f}^{\boldsymbol{\cN} \backslash S}, \hat{f}^S})$ is not equal to $\bf J^S(\mathbf{\hat{f}^{\boldsymbol{\cN} \backslash S}, \hat{f}^S})$, as the former is a cost function and the latter a vector of user costs. 
We now define our mapping of $V(S)$.
\begin{definition}
	\label{worst_case_coalitional}
	For every coalition $S \subseteq \cN$: 
	$$V(S) \triangleq  \{\mathbf{J^S}(\mathbf{\hat{f}^{\boldsymbol{\cN} \backslash S}, \hat{f}^S})\}.$$
\end{definition}
Thus, given a coalition of users $S$, $V(S)$ is a set of all possible cost vectors in which $\cN \backslash S$ sends its flow according to (\ref{eq:malicious_maximin_strategy}) and $S$ sends its flow according to (\ref{eq:coalition_maximin_strategy}). 
It is clear that $V(S)$ satisfies the conditions of an NTU coalitional game. Closedness follows directly from Assumption S2 and following \cite{BO12, Myerson}, it can be shown that 
$V(S)$ is convex.
Having defined the worst-case coalitional game, we proceed to investigate it through the study of several (fair and stable)
solution concepts of cooperative game theory.

\subsection{The core}
We consider the game as defined in Definition \ref{worst_case_coalitional} and continue to describe its {\em core}, i.e., a set of solutions 
that are stable against coalitional deviations \cite{Myerson}. 
\begin{definition}
\label{Core}
Given an NTU coalitional game $V(\cdot)$, the core of the game
is a set of 
cost vectors $\mathcal{J}_c \subseteq \cJ$ such that $\forall \mathbf{J} \in \mathcal{J}_c$, $\forall S \subseteq \cN$ and for any $\mathbf{\bar{J}} \in \mathcal{J}$:
\begin{itemize}
\item $\mathbf{J} \in V(N)$.
\item {\normalfont If} $\bar{J}^i < J^i,~\forall i \in S$, {\normalfont then} $\mathbf{\bar{J}^S} \notin V(S)$.
\end{itemize}
\end{definition}
Translated to our scenario, a feasible cost vector lies in the core if there does not exist any coalition $S$, such that all users in $S$ strictly decrease their cost when sending their demand according to any maximin flow profile $(\mathbf{\hat{f}^{\boldsymbol{\cN} \backslash S}, \hat{f}^S})$.
We continue with the following theorem.
\begin{theorem}
\label{thm:bottleneck_core}
Consider a bottleneck routing game. Any system-optimal flow profile corresponds to a cost vector that lies in the core. 
\end{theorem}
\begin{proof}
According to Lemma \ref{lem:larger_demand_smaller_cost}, for any two coalitions
$S,T \subseteq \cN$ with $r^{T} \geq r^{S}$ it follows that
$J^{T}(\mathbf{\hat{f}^{\boldsymbol{\cN} \backslash T}, \hat{f}^{T}}) \leq J^{S}(\mathbf{\hat{f}^{\boldsymbol{\cN} \backslash S}, \hat{f}^{S}}).$
In particular, for the grand coalition $\cN$ and any coalition $S \subseteq \cN$ it follows that
$$J^*_{sys} = J^{\cN}(\mathbf{\hat{f}^{\boldsymbol{\cN}}}) \leq J^{S}(\mathbf{\hat{f}^{\boldsymbol{\cN} \backslash S}, \hat{f}^{S}}).$$
Consequently, for any coalition $S \subseteq \cN$, 
\begin{align}\label{eq:opt_vs_maximin}
J^*_{sys} 
&\leq \{\max_{\{e\in E| \sum_{k \in S} \hat{f}^k_e>0\}} T_e(\hat{f}^{\cN \backslash S}_e + \hat{f}^S_e)\}
\\ \nonumber
&=\max_{k \in S} \{\max_{e\in E| \hat{f}^k_e>0} T_e(\hat{f}^{\cN \backslash S}_e + \hat{f}^S_e)\}.
\end{align}
Consider a system-optimal flow profile $\bf f^*$ and 
denote the flow of user $i$ on link $e$ as $f^{*i}_e$. 
From (\ref{eq:opt_vs_maximin}) it follows that for any $i \in \cN$ and any $S \subseteq \cN$, 
\begin{align}
\max_{e \in E|f^{*i}_e>0} T_e(f^*_e) &\leq \max_{e \in E|f^*_e>0} T_e(f^*_e)
\\ \nonumber
&\leq \max_{k \in S} \{\max_{e\in E| \hat{f}^k_e>0} T_e(\hat{f}^{\cN \backslash S}_e + \hat{f}^S_e)\}.
\end{align}
Hence, there exists some coalition $S$ and user $j \in S$ for which 
\begin{align*}
J^j(\mathbf{f^*})= \max_{e \in E|f^{*j}_e>0} T_e(f^*_e) 
&\leq \max_{e\in E| \hat{f}^j_e>0} T_e(\hat{f}^{\cN \backslash S}_e + \hat{f}^S_e) 
\\ 
&= J^j(\mathbf{\hat{f}^{\boldsymbol{\cN} \backslash S}, \hat{f}^{S}}).
\end{align*}
In other words, for any coalition $S \subseteq \cN$, any maximin flow profile $(\mathbf{\hat{f}^{\boldsymbol{\cN} \backslash S}, \hat{f}^S})$ and any system-optimal flow profile $\bf f^*$, there always exists some user $j \in S$ which receives a higher (or equal) cost at $(\mathbf{\hat{f}^{\boldsymbol{\cN} \backslash S}, \hat{f}^S})$ compared to $\bf f^*$.
Thus, any system-optimal flow profile $\bf f^*$ corresponds to a cost vector that lies in the core.
\end{proof}
In the proof of Theorem \ref{thm:bottleneck_core} we established that for any coalition $S \subseteq \cN$, there exists a user that prefers the system-optimal flow profile over any maximin flow profile. We will show that the same result holds with respect to the system-optimal flow profile and any Nash equilibrium. 

Suppose we consider a different coalitional game than the one described in Definition \ref{worst_case_coalitional}, in which a deviating coalition $S$ does not incur a maximin cost.
Specifically, in this new game the users in the coalition $S$ work together as a single large user, while the users outside the coalition do not aim to maximize the cost of $S$. Instead, they do not cooperate and their goal is to minimize their own costs. 
Denote the set of users outside the coalition as $\{1,2,\ldots, N_S\}$, where $N_S = |\cN \backslash S|$.
Thus, for each deviating coalition $S \subseteq \cN$, we consider a non-cooperative game between the users $\{S, 1, \ldots, N_S\}$, where each user $i$ aims to minimize its own cost. 
The next 
theorem is proven in Appendix \ref{app:NEP_coalitional_game}. 
\begin{theorem}\label{thm:NEP_coalitional_game}
Consider users with bottleneck objectives and consider an NTU coalitional game, where for any $S \subseteq \cN$, $V(S)$ is equal to the set of Nash equilibria between $S$ and all individual users in $\cN \backslash S$. 
Any system-optimal flow profile corresponds to a cost vector that lies in the core. 
\end{theorem}
\noindent For the remainder of the paper we only consider the worst-case NTU coalitional game as defined in Definition \ref{worst_case_coalitional}.

\subsection{The Routing Game Nucleolus}
Theorem \ref{thm:bottleneck_core} establishes that in a bottleneck routing game, no coalition of users wishes to deviate from any flow profile which is optimal from a system's perspective. 
Yet, 
out of all the stable flow profiles
in the core, which one should be proposed to the users in the network? 
We proceed to consider a further {\em refinement} of the core 
with additional fairness guarantees. 
In particular, we focus on a 
routing strategy flow profile which preserves stability while adding {\em min-max fairness} properties to the users. 
For that, we consider a variant of the well studied Nucleolus \cite{Schmeidler_Nuc}. 
Although the nucleolus is originally defined for Transferable-Utility games, we shall adapt its definition to fit our model. Due to this change, certain attributes such as its uniqueness are not guaranteed, unless proven. Moreover, we will need to prove that the nucleolus lies in the core of the game.
Consider a feasible cost vector $\mathbf{J(f)} \in \cJ$. 
We define the excess of each coalition $S \subset \cN$ at $\mathbf{J(f)}$ as
\begin{equation}\label{eq:excess}
e_S(\mathbf{J(f)}) \triangleq J^S(\mathbf{f}) - J^S(\mathbf{\hat{f}^{\boldsymbol{\cN} \backslash S}, \hat{f}^S}),
\end{equation}
where $J^S(\mathbf{f})$ corresponds to the coalition's cost at flow profile $\bf f$.
For bottleneck routing games, $J^S(\mathbf{f})$ corresponds to (\ref{eq:maximin_bottleneck_cost}), thus the excess of a coalition $S$ is equal to
\begin{align} \label{eq:bottleneck_excess}
&e_S(\mathbf{J(f)}) \triangleq \max_{e \in E| f^S_e>0} T_e(f_e) -  \max_{e \in E| \hat{f}^S_e>0} T_e(\hat{f}^{\cN \backslash S}_e + \hat{f}^S_e)
\\ \nonumber
&= \max_{i \in S} \{\max_{e \in E| f^i_e>0} T_e(f_e) \}-  \max_{i \in S} \{\max_{e \in E| \hat{f}^i_e>0} T_e(\hat{f}^{\cN \backslash S}_e + \hat{f}^S_e)\}.
\end{align}
For ease of notation we shall refer to $e_S(\mathbf{J(f)})$ as $e_S(\mathbf{J})$. 
The excess can be interpreted as the 
``dissatisfaction'' of a coalition $S$ at a cost vector $\bf J(f)$.
The smaller the excess, the more a coalition $S$ is satisfied. 
Note that even though the maximin flow profile is not necessarily unique, due to (\ref{eq:malicious_maximin_strategy}), $J^S(\mathbf{\hat{f}^{\boldsymbol{\cN} \backslash S},\hat{f}^S})$ is equal for all maximin flow profiles. 
Denote the excess vector $\mathbf{e}(\mathbf{J}) = [e_S(\mathbf{J})]_{S \subset N}$, and denote $\mathbf{e}^*(\mathbf{J})$ as a permutation of the entries of $\mathbf{e}(\mathbf{J})$ arranged in non-increasing order. 
\begin{definition}
\label{def:bottleneck_nucleolus}
We define the Routing Game Nucleolus (RGN) as:
$\{\mathbf{J} \in \cJ \mid \nexists \mathbf{\bar{J}} \in \cJ, e^*(\mathbf{\bar{J}}) \prec_{lxm} e^*(\mathbf{J}) \}$,
where $\prec_{lxm}$ means that it is smaller in the lexicographical sense.
\end{definition}
Hence, the RGN is the cost vector that minimizes the excess of the coalitions in the lexicographic ordering, i.e., it 
treats the 
welfare of coalitions in a min-max fair manner. 
We denote the vector of user costs at the Routing Game Nucleolus as $\bf J_{RGN}$.
\begin{proposition}\label{prop:bottleneck_nuc_core}
Consider a bottleneck routing game. The Routing Game Nucleolus lies in the core.
\end{proposition}
\begin{proof}
From Lemma \ref{lem:larger_demand_smaller_cost} it follows that for any coalition $S \subset \cN$,
$J^*_{sys} \leq J^S(\mathbf{\hat{f}^{\boldsymbol{\cN} \backslash S}, \hat{f}^S})$.
Therefore, due to (\ref{eq:excess}), the excess of 
any coalition $S \subseteq \cN$ at $\mathbf{J}^* \triangleq [J^*_{sys}]_{i \in \cN}$ is equal to 
\begin{equation}\label{eq:highest_excess}
e_S(\mathbf{J^*}) = J^*_{sys} - J^S(\mathbf{\hat{f}^{\boldsymbol{\cN} \backslash S}, \hat{f}^S}) \leq 0.
\end{equation}
Consequently, from Definition \ref{def:bottleneck_nucleolus} it follows that for any coalition $S \subset \cN$, $e_S(\mathbf{J_{RGN}}) \leq \max_{S \subseteq \cN}{e_S(\mathbf{J^*})} \leq 0$.
Denote the flow profile at the RGN as $\bf f$. 
It follows from (\ref{eq:bottleneck_excess}) that for any coalition $S \subset \cN$, 
\begin{equation}\label{eq:nucleolus_core}
\max_{i \in S} \{\max_{e \in E| f^i_e>0} T_e(f_e)\}
\leq
\max_{i \in S} \{\max_{e \in E| \hat{f}^i_e>0} T_e(\hat{f}^{\cN \backslash S}_e + \hat{f}^S_e)\}
.
\end{equation}
As a result of (\ref{eq:nucleolus_core}), for any coalition $S$ there exists some user $k \in S$ for which
\begin{align*}
J^k(\mathbf{\hat{f}^{\boldsymbol{\cN} \backslash S}, \hat{f}^S})
&= \max_{i \in S} \{\max_{e \in E| \hat{f}^i_e>0} T_e(\hat{f}^{\cN \backslash S}_e + \hat{f}^S_e)\}
\\
&\geq \max_{i \in S} \{\max_{e \in E| f^i_e>0} T_e(f_e) \} \geq J^k_{RGN}.
\end{align*}
In other words, for any coalition $S$ there exists some user $k \in S$ that receives a lower cost at the nucleolus compared to any maximin flow profile. Thus, the nucleolus lies in the core.
\end{proof}
We now continue to describe the RGN. 
Before doing so, we bring the following definition from \cite{BlocqO15}. 
\begin{definition}\label{def:balanced} 
	A flow profile $\bf f$ is referred to as {\em balanced} if for any two paths $p_1,p_2 \in \cP$ with $f_{p_1}>0$, it holds that
	$\max_{e \in p_1} T_e(f_e)  \leq \max_{e \in p_2} T_e(f_e)$.
\end{definition}
In other words, at a balanced flow profile, every possible path with positive flow has the same cost for its worst-case link. In \cite{BlocqO15}, the next lemma is proven. 
\begin{lemma}
	\label{lem:opt_balanced}
	{\normalfont \textbf{[Lemma 1 in \cite{BlocqO15}]}} Consider a bottleneck routing game.
	Any system-optimal flow profile is balanced. \qed
\end{lemma}
For the following theorem denote the user with the smallest demand as $j$, i.e., $r^j = \min_{k \in \cN} r^k$.
\begin{theorem} \label{thm:bottleneck_nucleolus}
Consider a bottleneck routing game.
\\ \textbf{(1):} 
$\mathbf{J_{RGN}} = [J^*_{sys}]_{i \in \cN}$, 
and/or 
\\
\textbf{(2):} $J_{RGN}^j > J^*_{sys}$, and
for all $k \in \cN \backslash j$, $J^k_{RGN} \leq J^*_{sys}$.  
\end{theorem}
\begin{proof}
Consider the coalition $K \triangleq \cN \backslash j$ and the cost vector $\mathbf{J}^* \triangleq [J^*_{sys}]_{i \in \cN}$. According to Lemma \ref{lem:larger_demand_smaller_cost}, for any coalition $S \subset \cN$,
\begin{equation}\label{eq:RGN_proof}
J^K(\mathbf{\hat{f}^{\boldsymbol{\cN} \backslash K}, \hat{f}^K}) \leq J^S(\mathbf{\hat{f}^{\boldsymbol{\cN} \backslash S}, \hat{f}^S}) 
~\text{and}~
e_{K}(\mathbf{J^*}) \geq e_{S}(\mathbf{J^*}).
\end{equation}
Denote the flow profile at the RGN as $\bf f$. 
Due to (\ref{eq:RGN_proof}) and Definition \ref{def:bottleneck_nucleolus}
it follows that $e_{K}(\mathbf{J_{RGN}}) \leq e_{K}(\mathbf{J^*})$. 
Hence,
\begin{equation}\label{eq:RGN_contra}
\max_{i \in {K}} \left\{\max_{e \in E| f^i_e>0} T_e(f_e) \right\} \leq  \max_{e \in E| f^*_e>0} T_e(f^*_e).
\end{equation}
Assume by contradiction that there exists some user $k \in K$ for which $J^k_{RGN} > J^*_{sys}$. 
Thus, 
$$\max_{e \in E| f^k_e>0} T_e(f_e) > \max_{e \in E| f^*_e>0} T_e(f^*_e),$$
which is a contradiction to (\ref{eq:RGN_contra}).
Hence, there does not exist any $k \neq j$ for which $J^k_{RGN} > J^*_{sys}$. 

Now suppose that $J^j_{RGN} < J^*_{sys}$ and for all $k \neq j$, $J^k_{RGN} \leq J^*_{sys}$. 
According to Lemma \ref{lem:opt_balanced}, any system-optimal flow profile is balanced.
Thus, every path with positive flow has the same cost for its worst-case link.
It follows that,
\begin{equation}\label{eq:balanced_system_cost}
\max_{e \in E|f^{*i}_e>0} T_e(f^*_e) = \max_{e \in E|f^*_e>0} T_e(f^*_e),~\forall i \in \cN
\end{equation}
i.e., for any $i \in \cN$, $J^i(\mathbf{f^*}) = J^*_{sys}$.
As a result, there does not exist a system-optimal flow profile $\bf f^*$ for which $J^j(\mathbf{f}^*) < J^*_{sys}$. Consequently, 
there must exist some user $k \neq j$ for which $J^k_{RGN} > J^*_{sys}$, which is a contradiction. 
Therefore, at the Routing Game Nucleolus, either \textbf{(1):} $J^i_{RGN} = J^*_{sys}$ for all $i \in \cN$, and/or \textbf{(2):} $J^j_{RGN} > J^*_{sys}$ and $J^k_{RGN} \leq J^*_{sys}$ for all $k \in \cN \backslash j$.
%
\end{proof}
Theorem \ref{thm:bottleneck_nucleolus} establishes that at the nucleolus of a bottleneck routing game, there are two possible scenarios. \textbf{(1)}: All users send their flow according to the system optimum and/or \textbf{(2)}: Only the user with the smallest demand sends its flow on the worst-case link in the system, while the other users receive a similar or better performance than at the system optimum. 

This result is quite counterintuitive, and combined with Lemma \ref{lem:opt_balanced}, Theorem \ref{thm:bottleneck_nucleolus} illustrates that the smallest user in the system is never better off at the nucleolus than at any system-optimal flow profile. 
In the following example, we illustrate two different networks, where each corresponds to one of the settings of Theorem \ref{thm:bottleneck_nucleolus}. 
\begin{example}
\begin{figure}[h!]
	\centering
	\includegraphics[width=0.3\textwidth]{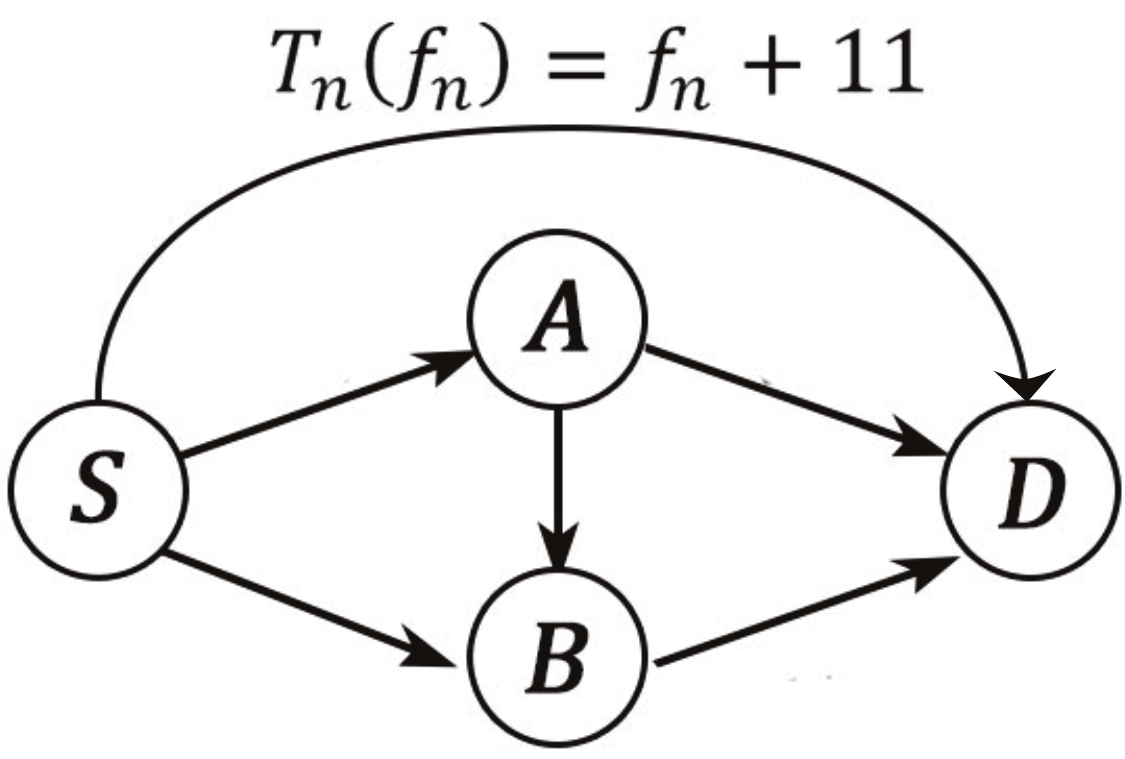}
	\caption{Nucleolus of a bottleneck routing game}
	\label{fig:example_2}
\end{figure}
Consider the network in Figure \ref{fig:example_2}, where for each link $e \in E \backslash \{n\}$, $T_e(f_e) = f_e$. Moreover, consider two users, $r^1=2$ and $r^2=20$. 
At the system-optimal flow profile, both users split their flow equally among the paths $\{S,A,D\}$ and $\{S,B,D\}$.
We first describe the maximin flow profile of user $2$. 
User $1$ maximizes the cost of user $2$ by placing all of its flow on the path $\{S,A,B,D\}$. As a best-response, 
user $2$ places $2/3$ on link $n$ and splits the rest of its flow 
equally among the paths $\{S,A,D\}$ and $\{S,B,D\}$. 
Thus, $J^{2}(\mathbf{\hat{f}^1, \hat{f}^2}) =11\frac{2}{3}$ and $e_{2}(\mathbf{J^*(f^*)}) = J^*_{sys}- J^{2}(\mathbf{\hat{f}^1, \hat{f}^2})= 11-11\frac{2}{3} = -\frac{2}{3}$. 
We now consider the excess of user $1$.
If user 2 places an amount of $5$ on link $n$ and sends the rest on path $\{S,A,B,D\}$, then the best-response strategy of users $1$ is to split its entire demand on the paths $\{S,A,D\}$ and $\{S,B,D\}$. Thus,  
$e_{1}(\mathbf{J^*(f^*)}) = J^*_{sys}- J^{1}(\mathbf{\hat{f}^2, \hat{f}^1}) \leq 11-16 = -5$.
However, if we propose a different flow profile, $\bf f$, where user $1$ places an amount of $1$ on link $n$ and splits the rest of its demand together with user 2 on the paths $\{S,A,D\}$ and $\{S,B,D\}$, we get that
$e_{2}(\mathbf{J(f)}) = 10.5- 11\frac{2}{3} = -1\frac{1}{6} \leq e_{2}(\mathbf{J^*(f^*)})$ and $e_{1}(\mathbf{J(f)}) \leq 12 - 16 = -4$.
Therefore, $\mathbf{J^*} \triangleq [J^*_{sys}]_{i \in \cN}$ is not equal to the RGN. 

Now consider a network as illustrated in Figure \ref{fig:example_2}, however without link $n$. It is straightforward that $\bf J^*$ is equal to the RGN.
\end{example}
We continue to focus on a special case in which there exist 
at least 
two ``smallest'' users in the system, i.e., there exist users $j_1, j_2$ such that $r^{j_1} = r^{j_2} = \min_{k \in \cN} r^k$. 
\begin{theorem}\label{thm:bottleneck__smallest_symmetric_nuc}
Consider a bottleneck routing game where there exist users $j_1, j_2$ such that $r^{j_1} = r^{j_2} = \min_{k \in \cN} r^k$. The Routing Game Nucleolus is equal to $[J^*_{sys}]_{i \in \cN}$.
\end{theorem}
\begin{proof}
Denote the grand coalition without user $j_1$ as $K_1 \triangleq \cN \backslash j_1$
and the grand coalition without user $j_2$ as $K_2 \triangleq \cN \backslash j_2$.
Clearly, $r^{K_1} = r^{K_2} \geq r^S$ for any coalition $S \subset \cN$. As a result of Lemma \ref{lem:larger_demand_smaller_cost} for any $S \subset \cN$,
\begin{equation*} 
J^{K_1}(\mathbf{\hat{f}^{\boldsymbol{\cN} \backslash K_1}, \hat{f}^{K_1}}) = J^{K_2}(\mathbf{\hat{f}^{\boldsymbol{\cN} \backslash K_2}, \hat{f}^{K_2}}) \leq J^{S}(\mathbf{\hat{f}^{\boldsymbol{\cN} \backslash S}, \hat{f}^{S}}).
\end{equation*}
Therefore, the excess of $K_1$ and $K_2$ at $\mathbf{J}^* \triangleq [J^*_{sys}]_{i \in \cN}$ is equal to 
\begin{align} \nonumber
e_{K_1}(\mathbf{J^*}) = e_{K_2}(\mathbf{J^*}) 
&= \max_{e \in E| f^*_e>0} T_e(f^*_e) - J^{K_1}(\mathbf{\hat{f}^{\boldsymbol{\cN} \backslash {K_1}}, \hat{f}^{K_1}})
\\ \label{eq:larger_excess}
&\geq e_{S}(\mathbf{J^*}),~\forall S \subset \cN.
\end{align}
Moreover, from Definition \ref{def:bottleneck_nucleolus} and (\ref{eq:larger_excess}), it follows that
\begin{equation}\label{eq:excess_nuc_smaller}
e_{K_1}(\mathbf{J_{RGN}}) \leq e_{K_1}(\mathbf{J^*}) ~\text{and}~ e_{K_2}(\mathbf{J_{RGN}}) \leq e_{K_2}(\mathbf{J^*}).
\end{equation}
Denote the flow profile at the RGN as $\bf f$ and assume by contradiction that there exists a user $k$ for which $J^k_{RGN} > J^*_{sys}$. 
Since user $k$ lies in $K_1$ or $K_2$ (or both) it holds that:
\begin{align*}
&\max_{i \in K_1} \left\{\max_{e \in E| f^i_e>0} T_e(f_e) \right\} > J^*_{sys}
\\
\mbox{and/or} ~&\max_{i \in K_2} \left\{\max_{e \in E| f^i_e>0} T_e(f_e) \right\} > J^*_{sys}.
\end{align*}
Therefore, $e_{K_1}(\mathbf{J_{RGN}}) > e_{K_1}(\mathbf{J^*})$ and/or $e_{K_2}(\mathbf{J_{RGN}}) > e_{K_2}(\mathbf{J^*})$, which is a contradiction to (\ref{eq:excess_nuc_smaller}).

Now assume by contradiction that there exists a user $k$ for which $J^k_{RGN} < J^*_{sys}$. 
According to Lemma \ref{lem:opt_balanced}, any system-optimal flow profile $\bf f^*$ is balanced. Thus, there does not exist a system-optimal flow profile $\bf f^*$ and a user $k$ for which $J^k(\mathbf{f}^*) < J^*_{sys}$. It follows that at the RGN there must exist another user $i$ for which $J^i_{RGN} > J^*_{sys}$, which is a contradiction. 
\end{proof}
The conditions of Theorem \ref{thm:bottleneck__smallest_symmetric_nuc} describe a range of scenarios for which the nucleolus is always equal to the system-optimal flow profile. 
For example, in a network where users have a finite number of different demands, e.g., ``large'', ``medium'' and ``small'', the conditions of Theorem \ref{thm:bottleneck__smallest_symmetric_nuc} are satisfied for a network with at least two ``small'' users. 
Of course Theorem \ref{thm:bottleneck__smallest_symmetric_nuc} also holds in a network where all users have equal demands, i.e., for all $i,j \in \cN$, $r^i = r^j$. We denote such users as symmetrical. 
\begin{corollary}\label{thm:bottleneck_symmetric_nuc}
Consider a bottleneck routing game with symmetrical users.  The Routing Game Nucleolus is equal to $[J^*_{sys}]_{i \in \cN}$.
\end{corollary}
Through Theorem \ref{thm:bottleneck_core} we established that a flow profile which is optimal from a system's perspective is also stable against coalitional deviations. Moreover, Proposition \ref{prop:bottleneck_nuc_core} and Theorem \ref{thm:bottleneck_nucleolus} establish that the min-max fair nucleolus is 
\textbf{(i)} stable and \textbf{(ii)} system-optimal or disadvantageous for the smallest user in the system. 
Finally, Theorem \ref{thm:bottleneck__smallest_symmetric_nuc} describes an interesting set of scenarios where the RGN is a desirable solution, from a fairness perspective, from a social perspective, and easy to implement due to its stability.  

\section{Additive Routing Games}\label{sec:additive}
\subsection{Maximin respresentation}
In contrast to bottleneck routing games, we now focus on users that are interested in additive performance measures, such as delay or packet loss. Moreover, we focus on the parallel links model as described in Section \ref{sec:model}. 
Thus, from \cite{ORS93}, $\bf f^*$ is a unique vector of link flows and the system-optimal proportional flow profile (\ref{proportional_alloc}) is unique in the individual link flows. 
Moreover, since a coalition $S$ behaves as a single user controlling the flow of its participants, it follows from (\ref{eq:additive_cost}) that
\begin{equation}
\label{eq:cost_additive_maximin}
J^S(\mathbf{f}) =  \sum_{i \in S}J^i(\mathbf{f}) = \sum_{l \in \cL} f^S_l T_l(f_l).
\end{equation}
Consequently, as stated in (\ref{eq:malicious_maximin_strategy}),
the maximin strategy of the users in $M \triangleq \cN \backslash S$ is equal to 
\begin{equation}
\label{eq:additive_maximin_strategy}
\mathbf{\hat{f}^M} = \argmax_{\bf f^M \in F^M} \{\min_{\bf f^S \in F^S}  \sum_{l \in \cL} f_l^S T_l(f^{M}_l+ f_l^S)\}
\end{equation}
and given $\mathbf{\hat{f}^M}$, the coalition $S$ send its demand according to 
\begin{equation}
\label{eq:additive_coalition_strategy}
\mathbf{\hat{f}^S}(\mathbf{\hat{f}^M}) = \argmin_{\bf f^S \in F^S}  \sum_{l \in \cL} f_l^S T_l(\hat{f}^{M}_l+ f_l^S).
\end{equation}
The (malicious) user $M$ aims to send its demand according to (\ref{eq:additive_maximin_strategy}) and 
sends an amount of flow $\hat{f}_l^M$ on each link $l \in \cL$, such that, when $S$ responds according to (\ref{eq:additive_coalition_strategy}), its cost is maximized.
To describe the behavior of 
$M$ at the maximin flow profile, we first establish several lemmas.
In each lemma we start out with a flow profile $\bf f = (f^M, \hat{f}^S(f^M))$, where $\bf \hat{f}^S(f^M)$ denotes the best-response routing strategy of $S$ given $\bf f^M$, i.e., given $\bf f^M$ the coalition $S$ sends its demand according to (\ref{eq:best-response-strategy}). We then change the strategy of the malicious user to $\bf g^M$, whereafter the coalition $S$ responds with a new best-response strategy
$\bf \hat{g}^S(g^M)$. Finally, we draw conclusions on the cost of $S$ at $\bf (g^M, \hat{g}^S(g^M))$. 
Denote the best-response strategy of $S$ to $\bf f^M$ and $\bf g^M$ as, respectively $\bf \hat{f}^S$, and $\bf \hat{g}^S$.
Consider the flow profile $\bf f = (f^M, \hat{f}^S)$ and order the links such that 
$T_1(f^M_1+\hat{f}^S_1) \leq \ldots \leq T_L(f^M_L + \hat{f}^S_L)$.
We define the set of the links with the lowest cost per unit of flow at $\bf f$ as
$$\cL^\sharp \triangleq \{l \in \cL \mid T_l(f^M_l + \hat{f}^S_l) = \min_{n \in \cL} T_n(f^M_n + \hat{f}^S_n)\}.$$
The first lemma establishes that when $M$ decreases the amount of flow it sends on the links with the lowest cost, it also lowers the cost of $S$.
\begin{lemma}
\label{lem:transfer_worst_better}
Consider a flow profile $\bf f = (f^M, \hat{f}^S)$, the set of links $\cL^\sharp$ and link $q \notin \cL^\sharp$. 
Moreover, consider a new routing strategy $\bf g^M$ and an amount of flow $\Delta \triangleq \sum_{l \in \cL^\sharp} \Delta_l >0$ such that for any link $l \in \cL^\sharp$ $g^M_l=f^M_l-\Delta_l$,
$g^M_q=f^M_q + \Delta$
and $\forall l \in \cL \backslash \{\cL^\sharp \cup q\},~ g^M_l=f^M_l$. 
Then 
\begin{equation}
\label{eq:transfer_wb}
J^S(\mathbf{g^M, \hat{g}^S}) < J^S(\mathbf{f^M, \hat{f}^S}).
\end{equation}
\end{lemma}
\begin{proof}
In order to establish the lemma, we construct an alternative 
routing strategy for $S$, $\bf \bar{g}^S$.  
By definition, the cost of $S$ when applying $\bf \hat{g}^S$ will be less or equal to its cost at $\bf \bar{g}^S$, i.e.,
$J^S(\mathbf{g^M,\hat{g}^S}) \leq J^S(\mathbf{g^M,\bar{g}^S})$.
Thus, it is sufficient to construct a feasible routing strategy $\bf \bar{g}^S$, which satisfies
\begin{equation}
\label{new_flow}
J^S(\mathbf{g^M,\bar{g}^S}) < J^S(\mathbf{f^M,\hat{f}^S}).
\end{equation}
We consider two cases:
\begin{enumerate}
\item $\Delta >\sum_{l \notin \cL^\sharp} \hat{f}^S_l$
\item $\Delta \leq \sum_{l \notin \cL^\sharp} \hat{f}^S_l$.
\end{enumerate}
Consider Case 1. We construct $\bf \bar{g}^S$ by sending all of $r^{S}$ randomly on the links $l \in \cL^\sharp$, while abiding by the constraint that $\bar{g}^S_l \leq \Delta_l + \hat{f}^S_l$ for all $l \in \cL^\sharp$. According to the conditions of Case 1, $\Delta > r^{S} - \sum_{l \in \cL^\sharp} \hat{f}^S_l$. Thus,
$$r^S = \sum_{l \in \cL^\sharp} \bar{g}^S_l < \sum_{l \in \cL^\sharp} [\Delta_l + \hat{f}^S_l].$$
Consequently, there exists some link $r$ for which $\bar{g}^S_r <\Delta_r + \hat{f}^S_r$.
Therefore, it follows that
\begin{align}\label{eq:subcase_1}
J^S(\mathbf{g^M,\bar{g}^S}) 
&= \sum_{l \in \cL^\sharp} \bar{g}^S_l T_l(f^M_l - \Delta_l + \bar{g}^S_l)
\\  \nonumber
&< \sum_{l \in \cL^\sharp} \bar{g}^S_l T_l(f^M_l + \hat{f}^S_l)
\\ \nonumber
&\leq \sum_{l \in \cL} \hat{f}_l^ST_l(f^M_l+\hat{f}_l^S) 
= J^S(\mathbf{f^M,\hat{f}^S}),
\end{align}
where the last inequality of (\ref{eq:subcase_1}) is due to the fact that
$\forall l\in \cL^\sharp, \forall n \notin \cL^\sharp$, $T_l(f^M_l+\hat{f}^S_l) < T_n(f^M_n+\hat{f}^S_n)$.
For Case 2, we consider two subcases.
\begin{enumerate}
\item [2a.]$\Delta \leq \hat{f}_q^S$.
\item [2b.]$\Delta > \hat{f}_q^S$.
\end{enumerate}
Consider Subcase 2a. We start at $\bf \hat{f}^S$ and construct $\bf \bar{g}^S$ by sending $\Delta$ from link $q$ to all the links $l \in \cL^\sharp$ such that $\bar{g}^S_l = \Delta_l + \hat{f}^S_l$. It follows that for all $l \in \cL^\sharp$, $T_l(g^M_l+\bar{g}^S_l) = T_l(f^M_l+\hat{f}^S_l)$ and $T_q(g^M_q+\bar{g}^S_q) = T_q(f^M_q+\hat{f}^S_q)$.
Therefore,
\begin{align*} 
J^S(\mathbf{g^M,\bar{g}^S}) 
&= \sum_{l \in \cL^\sharp} (\hat{f}^S_l+\Delta_l) T_l(f^M_l+\hat{f}^S_l) 
\\ \nonumber
&+ (\hat{f}^S_q - \Delta)T_q(f^M_q+\hat{f}^S_q)
\\ \nonumber
&+\sum_{l \in \cL \backslash\{\cL^\sharp \cup q\}} \hat{f}_l^ST_l\left(f^M_l+ \hat{f}_l^S\right) 
\\ \nonumber
&< \sum_{l \in \cL} \hat{f}_l^ST_l(f^M_l+\hat{f}_l^S)
= J^S(\mathbf{f^M,\hat{f}^S}),
\end{align*}
since $\forall l\in \cL^\sharp$, $T_l(f^M_l+\hat{f}^S_l) < T_q(f^M_q+\hat{f}^S_q)$.

Now consider Subcase 2b.
We start at $\bf \hat{f}^S$ and construct $\bf \bar{g}^S$ by sending an amount of flow $\hat{f}_q^S$ from link $q$ to the links in $\cL^\sharp$ and randomly consider a different link $q' \notin \cL^\sharp$. 
If $\Delta -  \hat{f}_q^S \leq \hat{f}_{q'}^S$ we finish constructing $\bf \bar{g}^S$ by sending $\Delta -  \hat{f}_q^S$ from link $q'$ to the links in $\cL^\sharp$. Otherwise we send $\hat{f}_{q'}^S$ from link $q'$ to the links in $\cL^\sharp$ and again consider a different link $q'' \notin \cL^\sharp$.
Through this process, we construct $\bf \bar{g}^S$ by sending a total amount of flow equal to $\Delta$ from a set of links with a higher cost per unit of flow than the links in $\cL^\sharp$.
Since $\cL$ is finite and 
$\Delta \leq \sum_{l \notin \cL^\sharp} \hat{f}^S_l$, we will reach a link $n \notin \cL^\sharp$ for which $\Delta - \sum_{k \in \cF} \hat{f}_k^S \leq \hat{f}_{n}^S$, where $\cF$ denotes the set of links we considered before reaching link $n$. 
We finish to construct $\bf \bar{g}^S$ by sending a total flow of $\Delta$ from the links $\{\cF \cup n\}$ to the links in $\cL^\sharp$.
Observe that for any $l \in \cL^\sharp$, 
\begin{align}\label{alg_pre_proof}
T_l(g^M_l+\bar{g}_l^S) &=T_l \left((f^M_l-\Delta_l)+(\hat{f}_l^S+\Delta_l)\right) 
\\ \nonumber
&= T_l(f^M_l+\hat{f}^S_l)
< T_r(f^M_r+\hat{f}^S_r),~\forall r \notin \cL^\sharp.
\end{align}
Consequently, from (\ref{alg_pre_proof}), 
\begin{align*}
J^S(\mathbf{g^M},\mathbf{\bar{g}^S}) &= 
\sum_{l \in \cL^\sharp} (\hat{f}_l^S+\Delta_l) T_l( f^M_l+\hat{f}_l^S)  
+ \bar{g}^S_n T_n(f^M_{n} + \bar{g}^S_n)
\\ 
&+ \sum_{l \in \cL \backslash\{\cL^\sharp \cup n \cup \cF\}} \hat{f}_l^ST_l(f^M_l+ \hat{f}_l^S) 
\\ 
&<  \sum_{l \in \cL} \hat{f}_l^ST_l(f^M_l+ \hat{f}_l^S) 
= J^S(\mathbf{f^M},\mathbf{\hat{f}^S}).
\end{align*}
Hence, given $\bf g^M$, we constructed a routing strategy $\bf \bar{g}^S$ that satisfies
(\ref{new_flow}). 
\end{proof}
We continue to establish that when $M$ increases the amount of flow it sends on the links with the lowest cost, it raises the cost of $S$.
\begin{lemma}
\label{lem:transfer_best_worst}
Consider a flow profile $\bf f = (f^M, \hat{f}^S)$, the set of links $\cL^\sharp$ and link $q \notin \cL^\sharp$. 
Moreover, consider a new routing strategy $\bf g^M$ and an amount of flow $\Delta \triangleq \sum_{l \in \cL^\sharp} \Delta_l >0$ such that for any link $l \in \cL^\sharp$ $g^M_l=f^M_l+\Delta_l$,
$g^M_q=f^M_q - \Delta$, $\forall l \in \cL \backslash \{\cL^\sharp \cup q\},~ g^M_l=f^M_l$, 
and
\begin{equation}\label{eq:link_constraint}
\left\{l \in \cL | T_l(g^M_l + \hat{g}^S_l) = \min_{n \in \cL} T_n(g^M_n + \hat{g}^S_n)\right\} = \cL^\sharp.
\end{equation}
Then $$J^S(\mathbf{g^M, \hat{g}^S}) > J^S(\mathbf{f^M, \hat{f}^S}).$$
\end{lemma}
\begin{proof}
The constraint in (\ref{eq:link_constraint}) implies that the set of links with minimal cost per unit of flow are equal in both $\bf (\mathbf{f^M, \hat{f}^S})$ and $\bf (\mathbf{g^M, \hat{g}^S})$.

Assume by contradiction that $J^S(\mathbf{g^M, \hat{g}^S}) \leq J^S(\mathbf{f^M, \hat{f}^S})$.
Consider a new routing strategy $\bf h^M$ where for all $l \in \cL^\sharp$ $h^M_l =g^M_l - \Delta_l$, $h^M_q = g^M_q + \Delta$ and
$\forall l \backslash \{\cL^\sharp \cup q\},~ h^M_l=g^M_l$. 
Due to (\ref{eq:link_constraint}) and Lemma \ref{lem:transfer_worst_better}, it follows that $J^S(\mathbf{h^M, \hat{h}^S}) < J^S(\mathbf{f^M, \hat{f}^S})$.
However, $\mathbf{h^M} =\mathbf{f^M}$, 
which is a contradiction. 
\end{proof}
Through Lemma \ref{lem:transfer_best_worst}, we have proven that it is in the malicious user's interest to send more flow on the links with the lowest cost per unit of flow. 
We are now able to establish the following theorem. 
\begin{theorem}
\label{thm:stack_worst}
Consider an additive routing game and a coalition $S \subseteq \cN$. At the maximin flow profile $\cN \backslash S$ sends its flow according to (\ref{non_atomic_user_conditions}), i.e., $\cN \backslash S$ sends its flow according to the best-response behavior of a set of non-atomic users with an aggregated demand of $r^{\cN \backslash S}$.
\end{theorem}
\begin{proof}
Consider a maximin flow profile $\bf (\hat{f}^{\boldsymbol{\cN} \backslash S}, \hat{f}^S)$ and denote the set 
$$\hat{\cL} \triangleq \left\{l \in \cL | T_l(\hat{f}^{\cN \backslash S}_l + \hat{f}^S_l) = \min_{n \in \cL} T_n(\hat{f}^{\cN \backslash S}_n + \hat{f}^S_n).\right\}$$

Assume by contradiction that there exists a link $q \notin \hat{\cL}$, for which $\hat{f}^{\cN \backslash S}_q>0$. 
We construct a feasible routing strategy 
$\bf g^{\boldsymbol{\cN} \backslash S}$ at which the malicious user moves 
a small amount of flow
$\Delta$ from link $q$ to the links in $\hat{\cL}$, such that $0<\Delta < \hat{f}_q^{\cN \backslash S}$ and
\begin{equation*}
\{l \in \cL | T_l(\hat{f}^{\cN \backslash S}_l + \Delta_l+\hat{f}^S_l) = \min_{n \in \cL} T_n(\hat{f}^{\cN \backslash S}_n + \Delta_n + \hat{f}^S_n)\} = \hat{\cL}
\end{equation*}
,i.e., such that the set of links in $\hat{\cL}$ does not change.
According to Lemma \ref{lem:transfer_best_worst}, this increases the cost of $S$. Thus, 
$\bf \hat{f}^{\boldsymbol{\cN} \backslash S}$ 
does not maximize (\ref{eq:malicious_maximin_strategy}),
which is a contradiction.
Hence, at its optimal routing strategy, $\cN \backslash S$ will send its flow according to (\ref{non_atomic_user_conditions}). 
\end{proof}
According to Theorem \ref{thm:stack_worst}, for any coalition $S \subseteq \cN$ at its maximin strategy, the users in $\cN \backslash S$ act as if they were a set of self-optimizing non-atomic users. 
Moreover, $S$ represents a single finite user and sends its flow according to its best-response strategy (\ref{eq:additive_coalition_strategy}). 
This specific scenario has been investigated in \cite{RichmanS07, Wan12} among others. 
In \cite{RichmanS07} it has been established that, for additive routing games, the setting of self-optimizing non-atomic users together with a single finite user $S$, admits a unique equilibrium. Therefore, 
as a result of Theorem \ref{thm:stack_worst}, we conclude that the maximin profile is unique for any coalition.
Further, in \cite{Wan12} the following lemma is established.
\begin{lemma}\label{lem:cheng_theorem}
{\normalfont \textbf{[Theorem 5.2 in \cite{Wan12}]}} Consider an additive routing game. For any two coalitions $S,T \subseteq \cN$ with $r^{T} \geq r^{S}$:
\begin{equation}
\label{cheng_theorem}
\frac{1}{r^{T}}  J^{T}(\mathbf{\hat{f}^{\boldsymbol{\cN} \backslash T}, \hat{f}^{T}}) \leq \frac{1}{r^{S}} J^{S}(\mathbf{\hat{f}^{\boldsymbol{\cN} \backslash S}, \hat{f}^{S}}).
\end{equation}
\end{lemma}
In other words, Lemma \ref{lem:cheng_theorem} implies that the average cost of a deviating coalition decreases when its aggregated demand increases. Note that this result is similar to Lemma \ref{lem:larger_demand_smaller_cost} for bottleneck routing games, although in Lemma \ref{lem:cheng_theorem} the costs are averaged. 

\subsection{The core}\label{sec:coalitions}
We continue to describe the core of our coalitional game with additive costs.
In the following proposition we establish that in general, not every system-optimal allocation
lies in the core. This is in contrast to Theorem \ref{thm:bottleneck_core}, which considers bottleneck routing games.
\begin{proposition} \label{prop:core}
Consider an additive routing game and suppose there exists a coalition $S$ and a maximin flow profile $(\mathbf{\hat{f}^{\boldsymbol{\cN} \backslash S}, \hat{f}^S})$, which is not system-optimal.
There exists a system-optimal flow profile whose corresponding cost vector does not lie in the core. 
\end{proposition}
\begin{proof}
Our goal is to construct a system-optimal flow profile $\bf \bar{f}$ such that 
for some coalition $S \subseteq \cN$ with maximin flow profile $(\mathbf{\hat{f}^{\boldsymbol{\cN} \backslash S}, \hat{f}^S})$ it holds that, 
\begin{equation}\label{eq:notin_core}
J^i(\mathbf{\bar{f}}) > J^i(\mathbf{\hat{f}^{\boldsymbol{\cN} \backslash S}, \hat{f}^S}),~\forall i \in S.
\end{equation}

Consider a coalition $S \subseteq \cN$ and a maximin flow profile $(\mathbf{\hat{f}^{\boldsymbol{\cN} \backslash S}, \hat{f}^S}) \neq \bf f^*$
as described in the proposition. 
Also consider the two complementary sets: $\cL^+ = \{l \in \cL | \hat{f}_l^{\cN \backslash S} \geq f_l^*\}$ and $\cL^- = \{l \in \cL| \hat{f}_l^{\cN \backslash S} < f_l^*\}$. 
We construct a new routing strategy for $S$, $\bf \bar{f}^S$ in which it ``fills'' up the links in $\cL^-$ according to the system optimum, starting from link $L$ upwards. 
Since $\sum_{l \in \cL^+} f^*_l - \hat{f}_l^{\cN \backslash S} \leq 0$, it follows that
$\sum_{l \in \cL^-} f^*_l - \hat{f}_l^{\cN \backslash S} \geq r^S$.
Therefore, after the filling process, $S$ reaches a link, $K \in \cL^-$ for which $\bar{f}^S_{K} \leq f^*_K - \hat{f}^{\cN \backslash S}_K$ and  for any link $l>K,~l \in \cL^-, \bar{f}^S_{l} = f^*_l - \hat{f}^{\cN \backslash S}_l$.
At the new flow profile, $S$'s cost is equal to
\begin{align}
\label{opt_nash_1}  \nonumber
J^S(\mathbf{\hat{f}^{\boldsymbol{\cN} \backslash S}, \bar{f}^S}) 
&= \sum_{\substack{l > K \\ l \in \cL^-}} [f_l^*-\hat{f}_l^{\cN \backslash S} ]T_l(f_l^*) 
\\ \nonumber
&+ \bar{f}_{K}^{S}T_{K}(\bar{f}_K^S+\hat{f}^{{\cN \backslash S}}_{K}) 
\\
&> J^S(\mathbf{\hat{f}^{\boldsymbol{\cN} \backslash S}, \hat{f}^S}),
\end{align}
because $\bf \hat{f}^S$ is the best-response strategy to $\bf \hat{f}^{{\boldsymbol{\cN} \backslash S}}$ and from \cite{ORS93} it follows that 
the best-response strategy is unique.
We now change the routing strategy of $\cN \backslash S$ and construct a flow profile $\bf \bar{f} \triangleq (\bar{f}^{\boldsymbol{\cN} \backslash S}, \bar{f}^S)$ that is system-optimal and for which $J^S(\mathbf{\bar{f}^{\boldsymbol{\cN} \backslash S}, \bar{f}^S}) \geq J^S(\mathbf{\hat{f}^{\boldsymbol{\cN} \backslash S}, \bar{f}^S})$. 
By doing so we have constructed a feasible system-optimal flow profile for which $J^S(\mathbf{\bar{f}^{\boldsymbol{\cN} \backslash S}, \bar{f}^S}) > J^S(\mathbf{\hat{f}^{\boldsymbol{\cN} \backslash S}, \hat{f}^S})$.

We define the strategy $\bf \bar{f}^{\boldsymbol{\cN} \backslash S}$ as follows. On any link $l \in \cL$, ${\cN \backslash S}$ sends an amount $\bar{f}^{\cN \backslash S}_l$ such that $\bar{f}^{\cN \backslash S}_l + \bar{f}^S_l = f^*_l$. Since for any link $l$, $\bar{f}_l^S \leq f_l^*$, this new routing strategy is feasible.
Thus, on any link $l \in \cL^-,~l > K$, user ${\cN \backslash S}$ does not increase its flow, i.e., $\bar{f}^{\cN \backslash S}_l = \hat{f}^{\cN \backslash S}_l$. Furthermore, on link $K$, $\bar{f}^{\cN \backslash S}_K \geq \hat{f}^{\cN \backslash S}_K$.
As a result,  
\begin{align}
\label{opt_nash_2}
&J^S(\mathbf{\bar{f}^{\boldsymbol{\cN} \backslash S}, \bar{f}^S}) 
\\ \nonumber
&= \sum_{\substack{l > K \\ l \in \cL^-}} \left[f_l^*-\bar{f}_l^{\cN \backslash S} \right]T_l(f_l^*) + \bar{f}_{K}^{S}T_{K}(\bar{f}_K^S+\bar{f}^{{\cN \backslash S}}_{K}) 
\\ \nonumber
&\geq \sum_{\substack{l > K \\ l \in \cL^-}} \left[f_l^*-\hat{f}_l^{\cN \backslash S} \right]T_l(f_l^*) + \bar{f}_{K}^{S}T_{K}(\bar{f}_K^S+\hat{f}^{{\cN \backslash S}}_{K}).
\end{align}
Hence, from (\ref{opt_nash_1}) and (\ref{opt_nash_2}), $J^S(\mathbf{\bar{f}^{\boldsymbol{\cN} \backslash S}, \bar{f}^S}) \geq J^S(\mathbf{\hat{f}^{\boldsymbol{\cN} \backslash S}, \bar{f}^S}) > J^S(\mathbf{\hat{f}^{\boldsymbol{\cN} \backslash S}, \hat{f}^S})$.
Now that we have constructed a system-optimal flow profile $\bf \bar{f}$, we can decide on how to split its cost among the individual users. 
Specifically, consider the system-optimal cost vector where all users in $S$ send their flow proportionally to $\bar{f}^S_l$. Thus, for every user $i \in S$, $\bar{f}^i_l = \frac{r^i}{r^S}\bar{f}^S_l$ and the cost of every user $i \in S$ is equal to
$\bar{J}^{i} = \frac{r^i}{r^S} J^S(\mathbf{\bar{f}^{\boldsymbol{\cN} \backslash S}, \bar{f}^S})$.
Furthermore, denote by $\hat{J}^{i}$ the cost of user $i \in S$ when sending its flow proportionally to $(\mathbf{\hat{f}^{\boldsymbol{\cN} \backslash S}, \hat{f}^S})$, i.e.,
$\hat{J}^{i} = \frac{r^i}{r^S} J^S(\mathbf{\hat{f}^{\boldsymbol{\cN} \backslash S}, \hat{f}^S})$. 
Clearly, it is in every users interest to deviate from the proposed system-optimal cost vector and send their flow proportionally to the maximin flow profile, i.e., for all $i \in S$, $\bar{J}^{i} > \hat{J}^{i}$, which satisfies (\ref{eq:notin_core}). 
It follows from Definition \ref{Core} that the system-optimal profile $\bf \bar{f}$, at which users in $S$ proportionally split their flow, does not lie in the core. 
\end{proof}
According to Proposition \ref{prop:core}, for additive routing games, in general not every system-optimal cost allocation lies in the core. Nevertheless, we focus on a particular cost allocation that does always lie in the core. 
 \begin{theorem}
\label{thm:core}
Consider an additive routing game. The Proportional Cost Allocation lies in the core. 
\end{theorem}
\begin{proof}
Assume by contradiction that the 
system-optimal 
Proportional Cost Allocation (\ref{proportional_alloc}), does not lie in the core. Hence, there exists a coalition $S$ for which it is worthwhile to deviate from the proportional flow profile and incur its cost at the maximin flow profile $(\mathbf{\hat{f}^{\boldsymbol{\cN} \backslash S}, \hat{f}^S})$, as defined in Definition \ref{worst_case_coalitional}.
Therefore, it follows that for all users $i \in S$
\begin{equation}
\label{core_theorem}
\frac{r^i}{R}J_{sys}^*  > J^i(\mathbf{\hat{f}^{\boldsymbol{\cN} \backslash S}, \hat{f}^S}).
\end{equation}
However, it follows from Lemma \ref{lem:cheng_theorem} that $\forall S \subseteq \cN$:
\begin{equation}
\nonumber
\frac{1}{R} J^{\cN}(\mathbf{\hat{f}^{\cN}}) = \frac{1}{R}J_{sys}^*  \leq \frac{1}{r^S} J^{S}(\mathbf{\hat{f}^{\boldsymbol{\cN} \backslash S}, \hat{f}^{S}}),
\end{equation}
which turns into
\begin{equation}
\label{indiv_rat_2}
\frac{r^S}{R} J_{sys}^*  \leq J^{S}(\mathbf{\hat{f}^{\boldsymbol{\cN} \backslash S}, \hat{f}^{S}}),~\forall S \subseteq \cN.
\end{equation}
Since, $\frac{r^S}{R} J_{sys}^* = \sum_{i \in S} \frac{r^i}{R} J_{sys}^*$ and $J^{S}(\mathbf{\hat{f}^{\boldsymbol{\cN} \backslash S}, \hat{f}^{S}}) = \sum_{i \in S} J^{i}(\mathbf{\hat{f}^{\boldsymbol{\cN} \backslash S}, \hat{f}^{S}})$, 
inequality (\ref{indiv_rat_2}) implies that if any coalition $S \subseteq \cN$ deviates from the PCA, there must exist some user $i \in S$, for which
$\frac{r^i}{R} J_{sys}^* \leq J^{i}(\mathbf{\hat{f}^{\boldsymbol{\cN} \backslash S}, \hat{f}^{S}})$. 
This is a contradiction to (\ref{core_theorem}). 
Therefore, the 
Proportional Cost Allocation lies in the core. 
\end{proof}
Theorem \ref{thm:core} illustrates that the 
proportional flow profile is not only system-optimal, but also stable against coalitional deviations. Moreover, as a result of Proposition \ref{prop:core}, in an additive routing game stability is a specific property of only certain system-optimal flow profiles, in particular the proportional flow profile. 

Note that Theorem \ref{thm:core} and Proposition \ref{prop:core} consider users with standard cost functions. In Appendix \ref{app:example:no_core} we provide an example of a network where the users' cost functions are not standard.
In that example, users have different performance objectives, and 
the proportional flow profile
does not lie in the core. This illustrates that Theorem \ref{thm:core} heavily depends on the setting of the system. 
Nevertheless, Theorem \ref{thm:core} states that, for users with standard additive performance objectives, if all users send their demand according to the proportional routing strategy, then (i) the system is optimal and (ii) none of the $2^N-1$ possible coalitions
would benefit by deviating from this strategy. 
Thus, for such users the proportional flow profile is easy to implement and highly desirable from a ``social'' perspective.

\subsection{The Routing Game Nucleolus}
We now continue to describe the RGN of our additive routing game. 
Due to the cost function in (\ref{eq:cost_additive_maximin}), for additive routing games we define a slightly different excess for each coalition, which is normalized in the coalition's demand. Specifically, for each coalition $S \subset \cN$ and a cost vector $\bf J(f)$, 
\begin{equation}\label{eq:addititve_excess}
e_S(\mathbf{J(f)})\triangleq \frac{1}{r^S}\left[ J^S(\mathbf{f}) - J^S(\mathbf{\hat{f}^{\boldsymbol{\cN} \backslash S}, \hat{f}^S}) \right].
\end{equation}
Hence, we consider the ``average dissatisfaction'' of a coalition $S$ at a cost vector $\bf J(f)$. Similarly to bottleneck routing games, we first prove that the RGN lies in the core. 
\begin{proposition}
Consider an additive routing game. The Routing Game Nucleolus lies in the core.
\end{proposition}
\begin{proof}
Denote the Proportional Cost Allocation as $\bf J^*$. 
From Lemma \ref{lem:cheng_theorem} it follows that for any coalition $S \subset \cN$, 
$\frac{r^S}{R}J^*_{sys} \leq J^S(\mathbf{\hat{f}^{\boldsymbol{\cN} \backslash S}, \hat{f}^S})$.
Therefore, due to (\ref{eq:addititve_excess}), the excess of 
any coalition $S \subseteq \cN$ at $\mathbf{J}^*$ equals
\begin{equation}\label{eq:additive_highest_excess}
e_S(\mathbf{J^*}) = \frac{1}{R}J^*_{sys} - \frac{1}{r^S}J^S(\mathbf{\hat{f}^{\boldsymbol{\cN} \backslash S}, \hat{f}^S}) \leq 0.
\end{equation}
Consequently, from Definition \ref{def:bottleneck_nucleolus} it follows that for any coalition $S \subset \cN$, $e_S(\mathbf{J_{RGN}}) \leq \max_{S \subseteq \cN}{e_S(\mathbf{J^*})} \leq 0$.
Denote the flow profile at the RGN as $\bf f$. 
It follows from (\ref{eq:addititve_excess}) that for any coalition $S \subset \cN$, 
$$\frac{1}{r^S}\sum_{i \in S}\sum_{l \in \cL} f^i_lT_l(f_l) 
\leq 
\frac{1}{r^S}\sum_{i \in S}\sum_{l \in \cL} \hat{f}^i_lT_l(\hat{f}^{\cN \backslash S}_l + \hat{f}^S_l).$$
Hence, for any coalition $S$ there exist a user $k \in S$ for which 
\begin{align*}
J^k_{RGN} =
\sum_{l \in \cL} f^k_lT_l(f_l) 
&\leq 
\sum_{l \in \cL} \hat{f}^k_lT_l(\hat{f}^{\cN \backslash S}_l + \hat{f}^S_l)
\\
&=J^k(\mathbf{\hat{f}^{\boldsymbol{\cN} \backslash S}, \hat{f}^S}).
\end{align*}
Consequently, the nucleolus lies in the core.
\end{proof}
We now continue to describe the nucleolus for symmetrical users.
\begin{theorem} \label{thm:nuc}
Consider an additive routing game with symmetrical users. The Routing Game Nucleolus 
is equal to the system-optimal Proportional Cost Allocation\footnote{
In Appendix \ref{app:nuc_not_PA} we provide an example of non-symmetrical users for which the PCA does not equal the RGN.}. 

\end{theorem}
\begin{proof}
Consider an additive routing game and denote the Proportional Cost Allocation as $\bf J^*$. 
Since all users have equal demands, for any two coalitions $S,T \subseteq \cN$ with $|T|>|S|$ it follows that $r^T > r^S$. Moreover, according to Lemma \ref{lem:cheng_theorem}, 
$\frac{1}{r^{T}} \hat{J}^{T}(\mathbf{\hat{f}^{\boldsymbol{\cN}\backslash {T}}, \hat{f}^{T}}) \leq \frac{1}{r^{S}} \ \hat{J}^{S}(\mathbf{\hat{f}^{\boldsymbol{\cN}\backslash {S}}, \hat{f}^{S}})$ and
$e_T(\mathbf{J^*}) \geq e_{S}(\mathbf{J^*})$. Similarly, if $|T|=|S|$ it follows that $e_T(\mathbf{J^*}) = e_{S}(\mathbf{J^*})$.
This implies that, at the PCA, any coalition of size $|S|=N-1$ has the largest excess. 
Altogether, there exist ${{N} \choose {N-1}} = N$ different combinations of coalitions for which $|S|=N-1$ and their excesses are equal. 

Now assume by contradiction that there exists some cost vector 
$\mathbf{\bar{J}}(\mathbf{\bar{f}}) \in \mathcal{J}$ 
for which $e^*(\mathbf{\bar{J}}) \prec_{lxm} e^*(\mathbf{J^*})$. 
Consider a coalition $K$ for which $|K| = N-1$. From Definition \ref{def:bottleneck_nucleolus} it follows that $e_S(\mathbf{\bar{J}}) \leq \max_{S\subseteq \cN}e_S(\mathbf{J^*}) = e_{K}(\mathbf{J^*})$. Therefore, it follows from (\ref{eq:addititve_excess}) that for any coalition $K$ with size 
$|S|=N-1$:
\begin{equation*} 
\frac{1}{r^{K}} \left[\bar{J}^K -
\hat{J}^K(\mathbf{\hat{f}^{\boldsymbol{\cN}\backslash K}, \hat{f}^K}) \right]
\leq \frac{1}{R} J^*_{sys} - \frac{1}{r^{K}} \hat{J}^K(\mathbf{\hat{f}^{\boldsymbol{\cN}\backslash K}, \hat{f}^K}).
\end{equation*}
Hence, for any coalition $K$ with size $|K|=N-1$:
\begin{equation}\label{eq:smaller_excess}
\sum_{i \in K} \bar{J}^i \leq  \frac{r^K}{R} J^*_{sys}.
\end{equation}
As a result of (\ref{eq:smaller_excess}), we get that 
\begin{equation}
\label{nucleolus_contra_3}
\sum_{\substack{K \subset N \\ |K|=N-1}} \sum_{i \in K} \bar{J}^i 
\leq 
\sum_{\substack{K \subset N \\ |K|=N-1}} \frac{r^K}{R}J_{sys}^*
=\sum_{\substack{K \subset N \\ |K|=N-1}} \frac{N-1}{N} J_{sys}^* .
\end{equation}
On the other hand we have
\begin{align}
\label{nuc_sym_2} 
\sum_{\substack{\forall K \subset N \\ |K|=N-1}}\sum_{i \in K} \bar{J}^i
=  (N-1)\sum_{i \in \cN} \bar{J}^i 
&\geq (N-1)J_{sys}^* 
\\ \nonumber
&= \sum_{\substack{\forall K \subset N \\ |K|=N-1}} \frac{N-1}{N}J_{sys}^*.
\end{align}
Therefore, by combining (\ref{nucleolus_contra_3}) with (\ref{nuc_sym_2}) it follows that 
$\sum_{i \in \cN} \bar{J}^i = J_{sys}^*$.
However, since $\bf \bar{J} \neq J^*$, there exists a user $k$ for which $\bar{J}^k <\frac{r^k}{R}J_{sys}^*$. Thus,
$$
\sum_{i \in \cN \backslash k} \bar{J}^i = J^*_{sys} - \bar{J}^k >
\frac{R-r^k}{R} J^*_{sys} = \frac{r^{\cN \backslash k}}{R} J^*_{sys},
$$
which is a contradiction to (\ref{eq:smaller_excess}). 
Therefore, there does not exist any $\bf \bar{J}\neq J^*$ for which $e^*(\mathbf{\bar{J}}) \prec_{lxm} e^*(\mathbf{J^*})$ and the Proportional Cost Allocation is equal to the 
Routing Game Nucleolus.
\end{proof}
Proposition \ref{prop:core} establishes that, in contrast to bottleneck games, not all system-optimal flow profiles are stable. 
Nevertheless, as a result of Theorem \ref{thm:core}, we established that the Proportional Cost Allocation is optimal from a system's perspective and is stable for users with additive cost functions.
Finally, Theorem \ref{thm:nuc} establishes that, for symmetric users, the PCA is equal to the nucleolus. 
Thus, it is highly recommendable from a network design perspective that 
users with additive performance objectives send their flow according to the proportional flow profile.
\section{Conclusions}
We investigated a coalitional routing game from a worst-case perspective.  
In particular, we described the cost of each of the $2^N$ coalitions as corresponding to their maximin flow profiles. This represents the performance that the coalition can guarantee itself, even under the pessimistic expectation that the agents outside the coalition have unpredictable or even malicious objectives. 
For bottleneck routing games, we established that any agreement that is optimal from a system's perspective, lies in the core. Such stability is highly desirable and makes 
it possible to implement the agreement (e.g., through a mediator), since no possible coalition of agents would benefit from deviating from the proposed solution.

Additionally, we established that the smallest agent in the system will never prefer the nucleolus over the system-optimal Proportional Cost Allocation.
Furthermore, we described an interesting set of scenarios at which a mediator, e.g., a network administrator, should propose to implement the system-optimal and min-max fair nucleolus. In particular, when the two smallest agents in the system are of equal size, the nucleolus corresponds to a system-optimal flow profile. 
On the other hand, in general the Nucleolus is hard to compute, (see, e.g., \cite{FaigleKK98} and references therein).
Thus, when the conditions of Theorem \ref{thm:bottleneck__smallest_symmetric_nuc} are not satisfied, due to considerations of complexity, a mediator might be better off proposing a stable agreement without min-max fairness guarantees such as the Proportional Cost Allocation. 

For agents with additive costs, our study focused on load balancing (routing) among servers (links). We establish that, at the maximin flow profile, the malicious users act as if they were a continuum of infinitesimal (i.e., {\em nonatomic}) self-optimizing agents. 
Using this result, we established that, in contrast to bottleneck routing games,
stability is generally not shared by every system-optimal allocation. 
Nevertheless, we showed that the system-optimal Proportional Cost Allocation lies in the core, and for symmetric agents it is equal to the nucleolus.
Hence, these results suggest that it would be advantageous for all agents (and the system) if they send their demand according to the system-optimal proportional flow profile.
Our results establish that such an agreement is stable, optimal from a system's perspective, and for symmetric agents it adds min-max fairness guarantees. 
Finally, we established that, when agents in the network have vastly different performance objectives, even the PCA does not guarantee stability. 
This suggests a design guideline that attempts to separate among homogeneous groups of users (e.g., “highly delay-sensitive”, “less delay-sensitive but highly sensitive to packet loss”, etc.) so that each group would agree to share its own network resources according to the Proportional Cost Allocation.



\bibliographystyle{IEEEtran}
\bibliography{worst_case_bib}

\begin{thebibliography}{10}
\providecommand{\url}[1]{#1}
\csname url@samestyle\endcsname
\providecommand{\newblock}{\relax}
\providecommand{\bibinfo}[2]{#2}
\providecommand{\BIBentrySTDinterwordspacing}{\spaceskip=0pt\relax}
\providecommand{\BIBentryALTinterwordstretchfactor}{4}
\providecommand{\BIBentryALTinterwordspacing}{\spaceskip=\fontdimen2\font plus
\BIBentryALTinterwordstretchfactor\fontdimen3\font minus
  \fontdimen4\font\relax}
\providecommand{\BIBforeignlanguage}[2]{{%
\expandafter\ifx\csname l@#1\endcsname\relax
\typeout{** WARNING: IEEEtran.bst: No hyphenation pattern has been}%
\typeout{** loaded for the language `#1'. Using the pattern for}%
\typeout{** the default language instead.}%
\else
\language=\csname l@#1\endcsname
\fi
#2}}
\providecommand{\BIBdecl}{\relax}
\BIBdecl

\bibitem{ORS93}
A.~Orda, R.~Rom, and N.~Shimkin, ``Competitive routing in multiuser
  communication networks,'' \emph{IEEE/ACM Trans. Netw.}, 1993.

\bibitem{altman_2000}
E.~Altman, T.~Basar, T.~Jim{\'{e}}nez, and N.~Shimkin, ``Competitive routing in
  networks with polynomial costs,'' \emph{{IEEE} Trans. Automat. Contr.},
  vol.~47, no.~1, pp. 92--96, 2002.

\bibitem{La02}
R.~J. La and V.~Anantharam, ``Optimal routing control: repeated game
  approach,'' \emph{IEEE Trans. Autom. Cont.}, vol.~47, pp. 437--450, 2002.

\bibitem{Roughgarden:2002}
T.~Roughgarden and E.~Tardos, ``How bad is selfish routing?'' \emph{J. {ACM}},
  vol.~49, no.~2, pp. 236--259, 2002.

\bibitem{KoutsoupiasP09}
E.~Koutsoupias and C.~H. Papadimitriou, ``Worst-case equilibria,''
  \emph{Computer Science Review}, vol.~3, no.~2, pp. 65--69, 2009.

\bibitem{YaicheMR00}
H.~Ya{\"{\i}}che, R.~Mazumdar, and C.~Rosenberg, ``A game theoretic framework
  for bandwidth allocation and pricing in broadband networks,''
  \emph{{IEEE/ACM} Trans. Netw.}, vol.~8, no.~5, pp. 667--678, 2000.

\bibitem{Han:2009}
Z.~Han and H.~V. Poor, ``Coalition games with cooperative transmission: a cure
  for the curse of boundary nodes in selfish packet-forwarding wireless
  networks,'' \emph{IEEE Transactions on Communications}, vol.~57, no.~1, 2009.

\bibitem{Gibbens:2008}
R.~J. Gibbens and P.~B. Key, ``Coalition games and resource allocation in
  ad-hoc networks,'' in \emph{Bio-Inspired Computing and Communication}.\hskip
  1em plus 0.5em minus 0.4em\relax Springer-Verlag, 2008, pp. 387--398.

\bibitem{LXWG2011}
D.~Li, Y.~Xu, X.~Wang, and M.~Guizani, ``Coalitional game theoretic approach
  for secondary spectrum access in cooperative cognitive radio networks,''
  \emph{{IEEE} Trans. Wireless Communications}, vol.~10, no.~3, pp. 844--856,
  2011.

\bibitem{SaadHDHB09}
W.~Saad, Z.~Han, M.~Debbah, A.~Hj{\o}rungnes, and T.~Basar, ``Coalitional games
  for distributed collaborative spectrum sensing in cognitive radio networks,''
  in \emph{Proceedings of INFOCOM}, 2009, pp. 2114--2122.

\bibitem{AntoniouKJPS09}
J.~Antoniou, I.~Z. Koukoutsidis, E.~Jaho, A.~Pitsillides, and I.~Stavrakakis,
  ``Access network synthesis game in next generation networks.'' \emph{Computer
  Networks}, vol.~53, pp. 2716--2726, 2009.

\bibitem{Tembine:2012}
X.~Luo and H.~Tembine, ``Evolutionary coalitional games for random access
  control,'' in \emph{Proceedings of {IEEE} {INFOCOM} 2013}, 2013, pp.
  535--539.

\bibitem{Singh:2012}
C.~Singh, S.~Sarkar, A.~Aram, and A.~Kumar, ``Cooperative profit sharing in
  coalition-based resource allocation in wireless networks,'' \emph{IEEE/ACM
  Trans. Netw.}, vol.~20, no.~1, pp. 69--83, 2012.

\bibitem{Saad_coalitionalgame}
W.~Saad, Z.~Han, M.~Debbah, A.~Hjorungnes, and T.~Basar, ``Coalitional game
  theory for communication networks,'' \emph{IEEE Signal Processing Magazine},
  vol.~26, no.~5, pp. 77--97, 2009.

\bibitem{Myerson}
R.~B. Myerson, \emph{Game theory - Analysis of Conflict}.\hskip 1em plus 0.5em
  minus 0.4em\relax Harvard University Press, 1997.

\bibitem{BO12}
G.~Blocq and A.~Orda, ``How good is bargained routing?'' \emph{IEEE/ACM
  Transactions on Networking}, vol.~PP, no.~99, pp. 1--15, 2016.

\bibitem{JNashNBS}
J.~Nash, ``The bargaining problem,'' \emph{Econometrica}, vol.~18, no.~2, pp.
  155--162, 1950.

\bibitem{Schmeidler_Nuc}
D.~Schmeidler, ``The nucleolus of a characteristic function game,'' \emph{SIAM
  Journal on Applied Mathematics}, vol.~17, no.~6, pp. 1163--1170, 1969.

\bibitem{Moscibroda:2006}
T.~Moscibroda, S.~Schmid, and R.~Wattenhofer, ``When selfish meets evil:
  byzantine players in a virus inoculation game,'' in \emph{Proceedings of
  {PODC} 2006}, 2006, pp. 35--44.

\bibitem{Karakostas:2007}
G.~Karakostas, T.~Kim, A.~Viglas, and H.~Xia, ``Selfish routing with oblivious
  users,'' in \emph{Proceedings of {SIROCCO} 2007}, 2007, pp. 318--327.

\bibitem{Chakrabarty}
D.~Chakrabarty, C.~Karande, and A.~Sangwan, ``The effect of malice on the
  social optimum in linear load balancing games,'' \emph{CoRR}, vol.
  abs/0910.2655, 2009.

\bibitem{KarakostasV07}
G.~Karakostas and A.~Viglas, ``Equilibria for networks with malicious users,''
  \emph{Math. Program.}, vol. 110, no.~3, pp. 591--613, 2007.

\bibitem{BabaioffKP09}
M.~Babaioff, R.~Kleinberg, and C.~H. Papadimitriou, ``Congestion games with
  malicious players,'' \emph{Games and Economic Behavior}, vol.~67, no.~1, pp.
  22--35, 2009.

\bibitem{altman:hal-01249188}
\BIBentryALTinterwordspacing
E.~Altman, A.~Singhal, C.~Touati, and J.~Li, ``{Resilience of Routing in
  Parallel Link Networks},'' {Inria Grenoble Rh{\^o}ne-Alpes, Universit{\'e} de
  Grenoble}, Research Report, 2015. [Online]. Available:
  \url{https://hal.inria.fr/hal-01249188}
\BIBentrySTDinterwordspacing

\bibitem{MaschlerGT}
M.~Maschler, E.~Solan, and S.~Zamir, \emph{Game Theory}.\hskip 1em plus 0.5em
  minus 0.4em\relax Cambridge University Press, 2013.

\bibitem{BannerO07}
R.~Banner and A.~Orda, ``Bottleneck routing games in communication networks,''
  \emph{{IEEE} Journal on Selected Areas in Communications}, vol.~25, no.~6,
  pp. 1173--1179, 2007.

\bibitem{BuschM09}
C.~Busch and M.~Magdon{-}Ismail, ``Atomic routing games on maximum
  congestion,'' \emph{Theor. Comput. Sci.}, vol. 410, no.~36, pp. 3337--3347,
  2009.

\bibitem{ColeDR12}
R.~Cole, Y.~Dodis, and T.~Roughgarden, ``Bottleneck links, variable demand, and
  the tragedy of the commons,'' \emph{Networks}, vol.~60, no.~3, pp. 194--203,
  2012.

\bibitem{HarksKM13}
T.~Harks, M.~Klimm, and R.~H. M{\"{o}}hring, ``Strong equilibria in games with
  the lexicographical improvement property,'' \emph{Int. J. Game Theory},
  vol.~42, no.~2, pp. 461--482, 2013.

\bibitem{Roughgarden04}
T.~Roughgarden, ``Stackelberg scheduling strategies,'' \emph{{SIAM} J.
  Comput.}, vol.~33, no.~2, pp. 332--350, 2004.

\bibitem{Nisan:2007}
N.~Nisan, T.~Roughgarden, E.~Tardos, and V.~V. Vazirani, \emph{Algorithmic Game
  Theory}.\hskip 1em plus 0.5em minus 0.4em\relax New York, NY, USA: Cambridge
  University Press, 2007.

\bibitem{Harks11}
T.~Harks, ``Stackelberg strategies and collusion in network games with
  splittable flow,'' \emph{Theory Comput. Syst.}, vol.~48, no.~4, pp. 781--802,
  2011.

\bibitem{rough_schopp11}
T.~Roughgarden and F.~Schoppmann, ``Local smoothness and the price of anarchy
  in atomic splittable congestion games,'' in \emph{Proceedings of {SODA}},
  2011, pp. 255--267.

\bibitem{Wan12}
C.~Wan, ``Coalitions in nonatomic network congestion games,'' \emph{Math. Oper.
  Res.}, vol.~37, no.~4, pp. 654--669, 2012.

\bibitem{KorilisLO97}
Y.~A. Korilis, A.~A. Lazar, and A.~Orda, ``Achieving network optima using
  stackelberg routing strategies,'' \emph{IEEE/ACM Trans. Netw.}, 1997.

\bibitem{Cominetti:2009}
R.~Cominetti, J.~R. Correa, and N.~E. Stier-Moses, ``The impact of
  oligopolistic competition in networks,'' \emph{Oper. Res.}, vol.~57, pp.
  1421--1437, November 2009.

\bibitem{WardropJ52}
J.~G. Wardrop, ``Some theoretical aspects of road traffic research,'' in
  \emph{Proceedings of the ICE, Pt. II}, vol.~1, 1952, pp. 325--378.

\bibitem{BO07}
R.~Banner and A.~Orda, ``Bottleneck routing games in communication networks,''
  \emph{IEEE Journal on Selected Areas in Communications}, vol.~25, no.~6, pp.
  1173--1179, 2007.

\bibitem{BlocqO15}
G.~Blocq and A.~Orda, ``"{B}eat-{Y}our-{R}ival" {R}outing {G}ames,'' in
  \emph{Proceedings of {SAGT}}, 2015, pp. 231--243.

\bibitem{RichmanS07}
O.~Richman and N.~Shimkin, ``Topological uniqueness of the nash equilibrium for
  selfish routing with atomic users,'' \emph{Math. Oper. Res.}, vol.~32, no.~1,
  pp. 215--232, 2007.

\bibitem{FaigleKK98}
U.~Faigle, W.~Kern, and J.~Kuipers, ``Computing the nucleolus of min-cost
  spanning tree games is np-hard,'' \emph{Int. J. Game Theory}, vol.~27, no.~3,
  pp. 443--450, 1998.

\bibitem{KKT}
H.~W. Kuhn and A.~W. Tucker, ``Nonlinear programming,'' in \emph{Proc. Second
  Berkeley Symp. on Math. Statist. and Prob}.\hskip 1em plus 0.5em minus
  0.4em\relax University of California Press, Berkeley, 1950, pp. 481--492.

\end{thebibliography}
\appendices
\renewcommand{\thetheorem}{\Alph{section}.\arabic{theorem}}
\renewcommand{\thelemma}{\Alph{section}.\arabic{lemma}}
\renewcommand{\thecorollary}{\Alph{section}.\arabic{corollary}}
\renewcommand{\theproposition}{\Alph{section}.\arabic{proposition}}
\renewcommand{\theclaim}{\Alph{section}.\arabic{claim}}
\renewcommand{\thedefinition}{\Alph{section}.\arabic{definition}}
\renewcommand{\theexample}{\Alph{section}.\arabic{example}}

\section{\small \textnormal{\textbf{Proof of Theorem \ref{thm:NEP_coalitional_game}}}}
\label{app:NEP_coalitional_game}
\begin{proof}
In order to prove the theorem we first investigate the Nash equilibria of a bottleneck routing game.
Consider a user $i \in \cN$ and denote the routing strategies of all users $\cN \backslash i$ as $\bf f^{-i}$. A flow profile $\bf f$ is a Nash equilibrium if for all $i \in \cN$,
\begin{equation}\label{eq:NEP}
J^i(\mathbf{f}) = \min_{\bf \tilde{f}^i \in F^i} J^i(\mathbf{\tilde{f}^i, f^{-i}}).
\end{equation}
In \cite{BO07} it is proven that such an equilibrium always exists for bottleneck routing games; in order to investigate its behavior we first provide the following definition as an extension of Definition \ref{def:balanced}.
\begin{definition}\label{def:balanced_in_flow}
	For any user $i \in \cN$, a flow profile $\bf f$ is referred to as {\em balanced} in $\bf f^i$, if for any two paths $p_1,p_2 \in \cP$ with $f^i_{p_1}>0$, it holds that
	$\max_{e \in p_1} T_e(f_e)  \leq \max_{e \in p_2} T_e(f_e)$.
\end{definition}
\noindent For the definition of a {\em balanced} flow profile we refer the reader to Definition \ref{def:balanced}. With the help of Definitions \ref{def:balanced} and \ref{def:balanced_in_flow}, we establish the following. 
\begin{lemma}\label{lem:NEP_balanced}
	Consider a bottleneck routing game.
	Any Nash equilibrium is balanced.
\end{lemma}
\begin{proof}
	Consider a Nash equilibrium $\bf f$ and a user $i \in \cN$. 
	For any fixed $\bf f^{-i}$, from the perspective of $i$, each link $e \in E$ 
	has an offset of $f^{-i}_e$. As a result, when $i$ sends its flow according to its best-response strategy $\bf f^i$, it sends its demand as a single optimizing user in a network with an offset of $f^{-i}_e$ on each link $e \in E$.
	Thus, according to Lemma \ref{lem:opt_balanced}, for any user $i \in \cN$, $\bf (f^i,f^{-i})$ is balanced in $\bf f^i$. 
	
	We now continue to prove that the Nash equilibrium $\bf f$ is balanced. 
	Assume by contradiction that  $\bf f$ is not balanced. Thus, there exist paths $p, p'$ with $f_p>0$, and 
	\begin{equation}\label{eq:not_balanced}
	\max_{e \in p} T_e(f_e) > \max_{e \in p'} T_e(f_e). 
	\end{equation}
	Therefore, there exists some user $j$ for which $f^j_p>0$. As a result of (\ref{eq:not_balanced}) and Definition \ref{def:balanced_in_flow}, $\bf f \triangleq (f^j, f^{-j})$ is not balanced in $\bf f^j$, which is a contradiction. 
	Hence, any Nash equilibrium flow profile $\bf f$ is balanced.
\end{proof}
Now consider a coalition $S \subseteq \cN$ and a bottleneck routing game with users $\cN_S \triangleq \{S, 1, \ldots, N_S\}$. Moreover, consider a Nash equilibrium $\bf \bar{f} = \{\bar{f}^S, \bar{f}^1, \ldots, \bar{f}^{N_S}\}$ and a system-optimal flow profile $\bf f^*$. 
According to Lemma \ref{lem:NEP_balanced}, $\bf \bar{f}$ is balanced. 
Thus, every path with positive flow has the same cost for its worst-case link and
$$\max_{e \in E|\bar{f}^i_e>0} T_e(\bar{f}_e) = \max_{e \in E|\bar{f}_e>0} T_e(\bar{f}_e),~\forall i \in \cN_S$$
i.e., for any $i \in \cN_S$, $J^i(\mathbf{\bar{f}}) = J_{sys}(\mathbf{\bar{f}})$. Moreover, as a result of Lemma \ref{lem:opt_balanced}, it follows that $\bf f^*$ is balanced. Thus, for any $i \in \cN_S$, 
\begin{equation}\label{eq:opt_system_indiv}
\max_{e \in E|f^{*i}_e>0} T_e(f^*_e) = \max_{e \in E|f^*_e>0} T_e(f^*_e),
\end{equation}
i.e., for any $i \in \cN_S$, $J^i(\mathbf{f^*}) = J_{sys}(\mathbf{f^*})$.
Consequently, 
\begin{equation}\label{eq:NEP_vs_optimum}
J^i(\mathbf{\bar{f}}) = J_{sys}(\mathbf{\bar{f}}) \geq J_{sys}(\mathbf{f^*}) = J^i(\mathbf{f^*}), ~\forall i \in \cN_S,
\end{equation}
where the inequality in (\ref{eq:NEP_vs_optimum}) follows from the definition of the system optimum. 
Inequality (\ref{eq:NEP_vs_optimum}) establishes that at any Nash equilibrium, all users are worse off than at any system-optimal profile. 
Hence, it is immediate that for any coalition $S \subseteq \cN$, any Nash equilibrium $\bf \bar{f}$ and any system-optimal flow profile $\bf f^*$, 
there exists a user $k \in S$ for which,
$J^k(\mathbf{\bar{f}})\geq J^k(\mathbf{f^*})$. Thus, any system-optimal flow profile $\bf f^*$ corresponds to a cost vector that lies in the core of the NTU game as described in Theorem \ref{thm:NEP_coalitional_game}. 
\end{proof}

\section{\small \textnormal{\textbf{ An additive routing game where users' cost functions are not standard and the PCA does not lie in the core.}}}\label{app:example:no_core}
Consider a network with two users, two parallel links and let the users' demands equal $r^1 = r^2 =0.5$.
The costs of the users are defined as follows:
\begin{equation*} 
J^1 = f^1_1  f_1 + 2 f_2^1 ( f_2 + 1)
,~J^2 = \frac{f_1^2}{1+\epsilon - f_1}+\frac{2f_2^2}{\epsilon^2} ( f_2 + 1),
\end{equation*}
where $0 < \epsilon < 0.1$. Note that, the user cost functions are not standard, since the performance function $T_e(\cdot)$ is different for both users.
We first investigate the maximin cost of user 2, 
i.e., $J^2(\mathbf{\hat{f}^1, \hat{f}^2})$.
For any given $\bf \hat{f}^1$, user $2$ sends its flow according to its best-response strategy (\ref{eq:best-response-strategy}). 
The minimization in (\ref{eq:best-response-strategy}) brings about the following necessary {\em Karush-Kuhn-Tucker (KKT)} \cite{KKT, ORS93} conditions: 
\begin{equation*} 
\text{if}~\hat{f}^2_2 > 0\rightarrow~
\frac{2(\hat{f}_2^1+\hat{f}^2_2+1)+2\hat{f}^2_2}{\epsilon^2}
\leq 
\frac{1+\epsilon-\hat{f}_1^1}{(1+\epsilon -\hat{f}^1_1-\hat{f}^2_1)^2}.
\end{equation*}
On the other hand,
\begin{equation*} 
\frac{1+\epsilon-\hat{f}_1^1}{(1+\epsilon -\hat{f}^1_1-\hat{f}^2_1)^2} 
\leq \frac{1+\epsilon}{\epsilon^2} 
< \frac{2}{\epsilon^2}
\leq 
\frac{2(\hat{f}_2^1+\hat{f}^2_2+1)+2\hat{f}^2_2}{\epsilon^2}.
\end{equation*}
It follows that for any given $\bf \hat{f}^1$, user 2 does not flow anything on the second link, i.e., $\hat{f}^2_2=0$.
Thus, $\hat{f}^2_1 = r^2 = 1/2$. Given $\hat{f}^2_1 = 1/2$, 
at the maxmin flow profile it is clear that user $1$ sends all of its demand on the first link, i.e., $\hat{f}^1_1 = r^1 = 1/2$.
Note that, for users without standard cost functions, the uniqueness of the system-optimal vector of link flows is not guaranteed \cite{ORS93}. Therefore, there might exist several proportional profiles, each corresponding to a different system-optimal strategy profile.
We need to establish that there does not exist any system-optimal proportional flow profile whose corresponding cost vector lies in the core.
Assume by contradiction that there exists a system-optimal proportional routing strategy $\bf f^*$ that lies in the core. 
At the proportional routing strategy, both users send their flow according to $f_l^i = \frac{f^*_l}{2}$ for $i=1,2$ and $\forall l \in \cL$. Thus, the costs of user 2 at 
$\bf f^*$ 
equals
\begin{equation*}
J^2(\mathbf{f}^*) = \frac{\frac{1}{2}f_1^*}{1+\epsilon - f_1^*}+\frac{f_2^*(f_2^*+1)}{\epsilon^2} \leq J^2(\mathbf{\hat{f}^1, \hat{f}^2}) = \frac{1}{2\epsilon},
\end{equation*}
where the inequality is due to the definition of the core.
This turns into
\begin{equation}\label{eq:costs_non_standard_6}
1 \geq \frac{\epsilon f_1^*}{1+\epsilon - f_1^*}+\frac{2(1-f_1^*)(2-f_1^*)}{\epsilon}.
\end{equation}
For any $0<\epsilon < 0.1$, (\ref{eq:costs_non_standard_6}) only holds if $f_1^* = 1$. Therefore, if any system-optimal proportional routing strategy $\bf f^*$ lies in the core, $f^*_1 = 1$ and $f^*_2=0$.
We have that $f^{1*}_1 = f^{2*}_1=\frac{1}{2}$, and
the social cost at $\bf f^*$ equals,
\begin{equation}\label{system_cost_non_standard}
\Jsys^* = \sum_{i \in \{1,2\}} J^i(\mathbf{f}^*) = \frac{1}{2\epsilon} + \frac{1}{2} \cdot 1 > 5\frac{1}{2}.
\end{equation}
Now consider a different flow profile, $\bf \bar{f}$, such that $f^2_1 = r^2 = 1/2$ and $f^1_2 = r^1 = 1/2$. The system cost equals 
\begin{equation}\label{system_cost_non_standard_2}
\bar{J}_{sys} =  \sum_{i \in \{1,2\}} J^i(\mathbf{\bar{f}}) = \frac{1/2}{1/2+\epsilon} + \frac{3}{2} < \frac{5}{2} < 5\frac{1}{2},
\end{equation}
which is a contradiction to $\bf f^*$ being optimal.

\section{\small \textnormal{\textbf{Example of non-symmetrical users for which the PCA does not equal the RGN.}}}\label{app:nuc_not_PA}
Consider a network with two users, two parallel links and let the users' demands equal $r^1 = 0.1$ and $r^2 =0.9$.
The costs of the users at $\bf f$ are defined as follows:
\begin{equation*}
J^1 = \frac{f_1^1}{2 - f_1}+\frac{f_2^1}{1 - f_2},~J^2 = \frac{f_1^2}{2 - f_1}+\frac{f_2^2}{1 - f_2}
\end{equation*}
At the maximin flow profile of any user, the other user places all of its demand on the first link. 
Denote the maximin profile when user $1$ deviates as $\bf (\hat{f}^2, \hat{f}^1)$ and the maximin profile when user $2$ deviates as $\bf (\tilde{f}^1, \tilde{f}^2)$.
Thus the cost of users $1$ and $2$ at $\bf (\hat{f}^2, \hat{f}^1)$ and $\bf (\tilde{f}^1, \tilde{f}^2)$ respectively equal:
\begin{align*}
J^1\mathbf{(\hat{f}^2, \hat{f}^1)} &= \frac{\hat{f}_1^1}{1.1 - \hat{f}_1^1}+\frac{\hat{f}_2^1}{1 - \hat{f}_2^1}
\\ \nonumber 
J^2\mathbf{(\tilde{f}^1, \tilde{f}^2)} &= \frac{\tilde{f}_1^2}{1.9 - \tilde{f}_1^2}+\frac{\tilde{f}_2^2}{1 - \tilde{f}_2^2}.
\end{align*}
By solving the necessary and sufficient KKT conditions \cite{KKT}, it follows that $\hat{f}^1_1 \approx 0.076$, $J^1\mathbf{(\hat{f}^2, \hat{f}^1)} \approx 0.099$, $\tilde{f}^2_1 \approx 0.74$, $J^2\mathbf{(\tilde{f}^1, \tilde{f}^2)} \approx 0.828$ and $J^*_{sys} \approx 0.9142$.
Denote the PCA as $\bf J^*$.
Thus, $e_{\{1\}}(\mathbf{J}^*) \approx -0.074$ and $e_{\{2\}}(\mathbf{J}^*) \approx -0.0062$.

Now consider a different optimal flow profile $\bf \bar{f}$ with cost vector $\bf \bar{J}$. At $\bf \bar{f}$, we let the users' flows equal $\bar{f}^1_1 = 0.075$, $\bar{f}^2_1 = 0.754$ and their costs equal $\bar{J}^1(\mathbf{\bar{f}}) \approx 0.094$, $\bar{J}^2(\mathbf{\bar{f}}) \approx 0.820$. 
It is easy to see that $e^*(\mathbf{\bar{J}(\bar{f})}) \prec_{lxm} e^*(\mathbf{J}^*)$. Hence, the proportional profile is not equal to the RGN.

\end{document}